\documentclass[preprint,12pt]{elsarticle}

\usepackage{amssymb}
\usepackage{amsmath}

\usepackage{fullpage}
\usepackage{bm}
\usepackage{color}
\usepackage{subcaption}
\usepackage[ruled,vlined]{algorithm2e}
\DeclareMathOperator*{\argmin}{arg\,min}
\DeclareMathOperator*{\argmax}{arg\,max}

\usepackage{amsthm}
\newtheorem{remark}{Remark}

\usepackage[hidelinks]{hyperref}

\journal{Journal of Computational Physics}

\begin{document}

\begin{frontmatter}



\title{The Level Set Ensemble Kalman Filter: Sequential Data Assimilation 
for Flows With Shocks}


\author[cal]{Michael K. Sleeman\corref{cor1}}
\ead{msleeman@caltech.edu}
\author[sd]{Lorenzo Beronilla}
\author[jh]{Hangchuan Hu}
\author[sd,cal]{\\Xu-Hui Zhou}
\author[sd]{Matthias Morzfeld}
\author[cal]{Andrew M. Stuart}
\author[jh]{Tamer A. Zaki}

\cortext[cor1]{Corresponding author.}

\affiliation[cal]{organization={Computing + Mathematical Sciences,
California Institute of Technology},
            city={Pasadena},
            postcode={91125}, 
            state={CA},
            country={USA}}
\affiliation[sd]{organization={Scripps Institution of Oceanography,
 University of California San Diego},
            city={La Jolla},
            postcode={92093}, 
            state={CA},
            country={USA}}
\affiliation[jh]{organization={Department of Mechanical Engineering,
Johns Hopkins University},
            city={Baltimore},
            postcode={21218}, 
            state={MD},
            country={USA}}

\begin{abstract}
The ensemble Kalman filter is commonly used to perform sequential
data assimilation in science and engineering because of its robust performance
and scalability when the state space dimension is high. However, the standard
EnKF generates spurious oscillations when applied to compressible flows with
shocks; the cause is the uncertainty, across the ensemble, in the shock
location. This paper develops the \emph{level set EnKF}, which overcomes this
challenge by employing a nonlinear mapping from the original state space into a
latent space in which the analysis step is undertaken. The latent
representation comprises a level set function that encodes the discontinuity
location and smooth state extensions, defined over the full physical domain,
that represent the solution on either side of the discontinuity. We first
formulate the level set EnKF as a specific instance of a more general approach
of performing data assimilation in a latent space. We then demonstrate the
feasibility of the level set EnKF for several one-dimensional compressible flows
and a two-dimensional blast wave.

\end{abstract}


\end{frontmatter}

\section{Introduction}

\subsection{Goal}\label{ssec:goal}

Sequential and variational data assimilation (DA) methods have proved successful
at combining predicted partial differential equation (PDE) models with partial,
noisy observations, to improve state estimates and enable uncertainty
quantification. However, many standard methods for large-scale problems fail in
the presence of shocks and other
discontinuities~\cite{houba2024da,edoh2025enkf,zhou2026neuralenkf}. In this
paper, we consider ensemble-based sequential DA for compressible flows with
sharp discontinuities (i.e., shocks and contact surfaces). We present a general
framework in which the analysis step is applied in a latent space, rather than
the original state space, to address the failure modes of standard ensemble
methods near discontinuities.

We then develop a specific implementation of this framework, which we term the
\textit{level set EnKF}. The method relies on three key ingredients:
(i) hyperbolic tangent profiles to model numerically smeared discontinuities,
(ii) gradient-based detection of candidate discontinuity locations, and
(iii) a nonlinear least-squares optimization to identify the
\textit{level set representation}. These three ingredients together result in a
latent\footnote{
The level set representation is higher dimensional than the physical space
representation, but a latent representation need not reduce the dimension of the
state; generally latent spaces are sought, in many areas of science and
engineering, to identify a space in which computations may be performed
favorably because, in some relevant topology, points are close together in the
latent space. The level set representation for shocks enables the EnKF analysis
update to preserve sharp features present in the forecast ensemble.
}
representation in terms of a level set function, which encodes discontinuity
location, and smooth state extensions, from either side of the discontinuity,
defined everywhere on the physical domain. We demonstrate that performing the
analysis step of EnKF in the resulting latent space preserves sharp
discontinuities. The approach is illustrated for several one-dimensional
(1D) compressible flows and for a two-dimensional (2D) blast wave problem.

\subsection{Literature Review and Context}

Physical phenomena can be investigated through a combination of observation
and mathematical modeling. Sensors provide physically grounded measurements
but are inherently limited by noise, finite spatio-temporal resolution, and
incomplete observability. In contrast, mathematical models based on PDEs, when
discretized and solved numerically, provide access to the complete state at
arbitrary resolution, but at the expense of errors, for example in the initial
and boundary conditions. DA offers a principled framework to combine models and
data to obtain improved state estimates~\cite{evensen2009data} and to perform
uncertainty quantification~\cite{reich2015probabilistic}. It has been applied to
diverse fields of science and engineering, including weather
forecasting~\cite{kalnay2002weather},
turbulence modeling~\cite{majda2012turbulence},
oil reservoir simulation~\cite{oliver2008inverse}
and other problems in the geophysical sciences~\cite{carassi2018geophysics}.
More recently, DA has been increasingly applied to flow estimation and
prediction in fluid mechanics~\cite{zhang2022ensemble,zhou2023inference,
zafar2026data,liu2025towards,wang2021state,zaki2025daTextbook,
zaki2025turbulence}. While much of this work has focused on incompressible and
smoothly varying flows, there is growing interest in DA for high-speed
compressible flows~\cite{buchta2021observation,buchta2022assimilation,
morra2024ml,wang2025domain} and for flows containing shocks and other
discontinuities~\cite{houba2024da,edoh2025enkf,zhou2026neuralenkf}.

We first review sequential DA methods for compressible flows and other
problems with discontinuities in Subsubsection~\ref{sssec:introFilter}. We then
review methods for shock detection in
Subsubsection~\ref{sssec:introShockDetection} and recent work on hyperbolic
tangent representations of discontinuities in
Subsubsection~\ref{sssec:introTanh}; these topics are relevant because the
proposed method requires that discontinuities be detected explicitly, and we
choose to model discontinuities using the hyperbolic tangent function.

\subsubsection{Sequential DA for Compressible Flows}\label{sssec:introFilter}

The \emph{filtering distribution} is defined through an iterative Bayesian
framework: model \emph{prediction} using the underlying PDE provides a
prior distribution; this prior information is updated using observational
data to obtain a posterior prediction in what is known as the \emph{analysis};
the process of prediction and analysis is cycled to incorporate observations
sequentially in time. Filtering algorithms, or simply filters, refer to a
class of sequential DA algorithms designed to approximate the filtering
distribution. Here we study filters for compressible flows modeled by the
compressible Euler equations (see \ref{app:eulerEquations}), whose solutions
contain shocks and contact surfaces.

The bootstrap particle filter (PF)~\cite{doucet2001introduction,
leeuwen_2009_BPF,arulampalam2002pf} employs a set of weighted
particles to approximate the
filtering distribution; it applies quite generally and, in particular, converges
to the true filtering distribution, in the large particle limit, without making
any Gaussian assumptions.  However, the method suffers from \emph{weight
collapse} in which all but one particle carry negligible weights. This issue is
particularly pronounced for problems in which the state space dimension is
high~\cite{snyder2008pf,snyder2015pf}, such as the spatially discretized Euler
equations. As a consequence, the ensemble Kalman filter
(EnKF)~\cite{evensen2009data,evensen1994sequential}, which works with equally
weighted ensemble members, is widely used in high dimensional settings.
However, the EnKF is most effective when uncertainties are well approximated by
Gaussian distributions. This assumption is problematic for compressible flows
with shocks: when the shock location is uncertain, the state distribution
at a fixed spatial location within the range of possible shock locations can
become bimodal, since different ensemble members may place that location on
different sides of the shock~\cite{zhou2026neuralenkf}. The EnKF analysis
step can then mix pre- and
post-shock states, producing nonphysical oscillations near the discontinuity, as
observed by Houba \textit{et al.}~\cite{houba2024da} for 1D shock propagation
problems.

Houba \textit{et al.}~\cite{houba2024da} show that the PF preserves shocks in
compressible flows, provided that all forecast states contain the same number of
shocks. However, weight collapse remains a problem in general. To mitigate
this issue, Subrahmanya and Sandu~\cite{subrahmanya2025etpfdtw} combine the
Ensemble Transform Particle Filter (ETPF)~\cite{reich2013etpf} with
\emph{dynamic time warping} (DTW)~\cite{sakoe1978dtw} to align features when
taking convex combinations of particles. They further employ underweighting
(i.e., inflation of the assumed observation-error covariance), to delay weight
collapse.

Despite this body of work, weight collapse remains a limitation of
particle-filter-based methods, motivating the development of
EnKF-based approaches for filtering problems with discontinuities. Recent
work by Edoh \textit{et al.}~\cite{edoh2025enkf} applied the EnKF to transformed
variables to enforce positivity, although this transformation does not address
the spurious oscillations caused by misaligned discontinuities. More recently,
several methods have instead applied the EnKF in a latent space.
Zhou \textit{et al.}~\cite{zhou2026neuralenkf} developed the
\textit{neural EnKF}, which represents the state using the weights and biases of
a neural network and performs DA in this latent space. Chandravamsi
\textit{et al.}~\cite{chandravamsi2026decoder} similarly learn a shared decoder
that maps latent variables to the physical state and perform DA in the learned
latent space. The \textit{morphing EnKF} of Beezley and
Mandel~\cite{beezley2008registration} instead constructs a latent space using
image registration~\cite{brown1992registration}, comprising an invertible
registration map together with the aligned state. Originally developed for
wildfire forecasting, it has recently been extended to compressible
flows~\cite{hu2026morphingenkf}.

The DA approaches above illustrate how performing the EnKF step in a latent
space yields an improved analysis step for fields with discontinuities. This
principle has also been explored for inverse problems by Chada
\textit{et al.}~\cite{chada2018levelset}, who used level set based latent
space to solve inverse problems with ensemble Kalman inversion (EKI). That paper
deploys the level set methodology to reconstruct piecewise constant  and
piecewise continuous fields. Their parameterization is related to the level set
representation that we develop in Section~\ref{sec:ls1d}.  Implementing the
resulting inversion methodology requires only a map from the latent space to
physical space, which is straightforward. In contrast, the present work
addresses a sequential filtering problem and therefore requires repeated
interaction with the PDE solver. Consequently, it also requires a map from
physical space to latent space, which is considerably more challenging. The
present work develops efficient tools for doing so.

\subsubsection{Shock Detection}\label{sssec:introShockDetection}

Shocks and contact surfaces have a finite width, but this width is small
compared to macroscopic flow scales, and they can be idealized to have
infinitesimal width~\cite{thompson1971compressible}. However, numerical
solutions to the compressible Euler equations smear shocks and
contact surfaces over multiple cells~\cite{toro2009compressible}, which makes
it challenging to precisely identify locations of discontinuities. Here, we
review methods for detecting shocks and contact surfaces.

\textit{Shock-capturing} methods solve hyperbolic conservation laws, such as the
Euler equations, while allowing shocks and other discontinuities to form without
explicitly tracking their locations. They use numerical dissipation to spread
discontinuities over a few grid cells and suppress spurious
oscillations~\cite{toro2009compressible}. This dissipation may be introduced
through artificial viscosity~\cite{neumann1950dissipation}, but is more commonly
built into Godunov-type schemes, including \textit{monotonic upstream-centered
schemes for conservation laws} (MUSCL)~\cite{vanleer1979muscl} and
\textit{weighted essentially non-oscillatory} (WENO)
methods~\cite{liu1994weno}. We refer to \ref{app:eulerEquations} and
Toro~\cite{toro2009compressible} for a more detailed treatment of
shock-capturing methods for the Euler equations.

Although shock-capturing methods do not explicitly track discontinuities, many
schemes that rely on artificial diffusion use \textit{shock sensors}\footnote{
Here, a \textit{shock sensor} refers to a numerical method used to identify
regions containing shocks or other sharp discontinuities, rather than a physical
measurement device.
}
to adaptively increase
artificial diffusion in regions with large solution gradients. Jameson
\textit{et al.}~\cite{jameson1981shocks} introduced an adaptive artificial
viscosity method based on a pressure sensor, Ducros
\textit{et al.}~\cite{ducros1999shockSensor} developed a related sensor based
on the velocity field, and Moro \textit{et al.}~\cite{moro2016shocks} later
proposed a dilation-based shock sensor. Despite their widespread use for
controlling artificial dissipation, shock sensors have been used less often for
explicit detection of shock location. We note, however, that Razavi and
Yano~\cite{razavi2025registration} used the dilation-based sensor of Moro
\textit{et al.} to locate shocks.

Wu \textit{et al.}~\cite{wu2013shockSensor} reviewed methods for
shock detection, including the \textit{Normal Mach number} approach of Lovely
and Haimes~\cite{lovely1999shockSensor} and the characteristic-based approach
of Kanamori and Suzuki~\cite{kanamori2011shockSensor}. More recent methods have
used image-processing and machine-learning techniques: Fujimoto
\textit{et al.}~\cite{fujimoto2019canny} used Canny edge
detection~\cite{canny1986edge}, Liu
\textit{et al.}~\cite{liu2019shockCNN} developed a convolutional neural network
(CNN)-based shock detector, and Chang
\textit{et al.}~\cite{chang2024shockCluster} developed a cluster-analysis method
to detect shocks. The proposed level set representation method (see
Section~\ref{sec:ls1d}) can be adapted to employ any future improvements in
robust shock detection.

\subsubsection{Modeling Sharp Gradients Using Hyperbolic Tangent Profiles}
\label{sssec:introTanh}

In the present work, shocks and contact surfaces are represented using
hyperbolic tangent profiles. This representation is related to the tangent of
hyperbola for interface capturing (THINC) method introduced by Xiao
\textit{et al.}~\cite{xiao2005thinc}, in which hyperbolic-tangent profiles are
used to capture moving interfaces in multi-fluid simulations. Baumgart
\textit{et al.}~\cite{baumgart2023tanh} similarly represented shocks using
hyperbolic tangent profiles and estimated the profile parameters through
least-squares regression; the present work employs a similar approach.

Although THINC was not originally developed to sharpen numerically smeared
discontinuities, Fukushima and Kitamura~\cite{fukushima2024sharp} applied it to
discontinuities smeared by MUSCL schemes. They found that the method sharpened
contact surfaces and weak shocks without introducing numerical oscillations,
whereas sharpening strong shocks produced spurious oscillations. We explore
using our hyperbolic tangent representation to sharpen shocks in
\ref{app:ls1dSharp}, and we find that doing so leads to oscillations. Therefore,
we sharpen only contact surfaces. More generally, numerical solutions of the
Euler equations introduce an intrinsic length scale associated with the width of
a shock, and we seek a level set representation that preserves this scale.

\subsection{Numerical Solutions of the Euler Equations}\label{ssec:introCFD}

In this work, we model compressible flows using the Euler equations, as
outlined in~\ref{app:eulerEquations}. We solve these equations numerically using
the open-source solver \texttt{M2C} (Multiphysics Modeling and
Computation)~\cite{zhao2026m2c}, which implements a MUSCL
scheme~\cite{vanleer1979muscl}. We compute numerical fluxes using a local
Lax-Friedrichs approximate Riemann solver and integrate in time using an
explicit second-order Runge-Kutta scheme. We refer to
Toro~\cite{toro2009compressible} for more details on compressible flow solvers.
Throughout this work, all level set representations and level set EnKF
updates are constructed from cell-centered solution values.

\subsection{Latent Space EnKF and the Level Set Representation}
\label{ssec:introLatentEnKF}

The neural-network-based approaches of Zhou
\textit{et al.}~\cite{zhou2026neuralenkf} and Chandravamsi
\textit{et al.}~\cite{chandravamsi2026decoder} perform the EnKF in a learned
latent space rather than directly in the physical state space. Similarly, the
morphing EnKF~\cite{beezley2008registration} defines a latent space consisting
of a registration map and the corresponding aligned state representation. In
this work, we present a general framework for latent-space EnKF methods, of
which these approaches are specific instances.

A suitable transformation to latent space should satisfy three properties:
\begin{enumerate}
    \item its inverse should accurately reconstruct the physical state;
    \item its inverse should be \textit{feature-preserving}; and
    \item the latent representation should vary smoothly across ensemble
    members.
\end{enumerate}
The first property limits reconstruction error, while the second ensures that
latent space perturbations preserve sharp physical features, such as shocks.
The third makes the latent space more amenable to the linear combinations
performed by the EnKF.

In the level set EnKF, level set functions encode discontinuity locations
through their zero level sets, while smooth state extensions represent the
solution on either side of each discontinuity. This representation is designed
to satisfy the criteria above: reconstructing from the latent variables
preserves sharp discontinuities, and separating discontinuity locations from
smooth state values yields latent variables that vary smoothly across ensemble
members.

\subsection{Contributions and Paper Outline}

The primary contribution of this paper is the development of the
\textit{level set representation} and its application to feature-preserving DA
for systems with discontinuities via the \textit{level set EnKF}. We show that
the method performs well for 1D and 2D compressible Euler problems with
uncertain shock locations. Beyond its application to DA, the level set
representation could also be used to extract shock properties from CFD data. In
particular, the representation yields the shock location and state values on
either side of the shock, allowing the shock strength and propagation speed to
be computed. Conversely, advances in shock-detection methods could improve
future implementations of the level set representation. A secondary contribution
is that we present a general framework for latent space EnKF; we place the level
set methodology in this framework and highlight that the existing
neural-network-based methods~\cite{zhou2026neuralenkf,chandravamsi2026decoder},
and the morphing EnKF~\cite{beezley2008registration}, are specific instances of
this framework.

Section~\ref{sec:transformedEnKF} introduces the notations for the standard and
transformed EnKFs. Section~\ref{sec:ls1d} develops the level set representation
for 1D scalar fields with one or more discontinuities, while
Section~\ref{sec:ls1dEuler} demonstrates its application to solve data
assimilation problems arising in the 1D Euler equations. In particular
Section~\ref{sec:enkf1dEuler} demonstrates the level set EnKF for the 1D
compressible Euler equations by applying it to Toro's shock tube
problem~\cite{toro2009compressible}, the Shu-Osher shock-entropy
problem~\cite{shu1989eno}, and Sod's shock tube problem~\cite{sod1978shockTube}.
Section~\ref{sec:ls2dEuler} extends the level set representation so that it may
be used to enable DA for solutions of the 2D Euler equations; and we illustrate
the methodology when applied to data assimilation for the blast wave problem
investigated in~\cite{subrahmanya2025etpfdtw}. We conclude in
Section~\ref{sec:conclusion}.

\section{EnKF and Discontinuous Flow Fields}\label{sec:transformedEnKF}

We first review the standard EnKF in Subsection~\ref{ssec:senkf}. We then
introduce the transformed EnKF in Subsection~\ref{ssec:tenkf}, discuss methods
to construct appropriate latent space transformation in
Subsection~\ref{ssec:tconstruct}, and demonstrate the transformed EnKF
for a hyperbolic tangent profile in Subsection~\ref{ssec:parametrizedHeaviside}.

\subsection{Standard EnKF} \label{ssec:senkf}

Consider a discretized physical state $\bm{z}_j\in\mathbb{R}^{N_z}$ at
observation times $t_j$ for $j=1,\dots,J$~\footnote{For ease of exposition we
assume that these times are equally spaced, but this assumption is readily
dispensed with in what follows.}. Let operator
$\mathcal{P}:\mathbb{R}^{N_z}\to\mathbb{R}^{N_z}$ advance the state between
observation times, assuming that this is defined by a (discretized) autonomous
PDE. This leads to the \emph{prediction} (forecast)
\begin{equation}
    \widehat{\bm{z}}_{j+1} = \mathcal{P}(\bm{z}_j),\quad j=1,\dots,J-1.
    \label{eq:dynamics}
\end{equation}
Here, and in what follows, $\widehat{(\cdot)}$ denotes a predicted quantity.

Let $\mathcal{H}:\mathbb{R}^{N_z}\to\mathbb{R}^{N_d}$ denote the observation
operator, which maps the state to the observation space. If both the dynamical
model~\eqref{eq:dynamics}, and the observation operator $\mathcal{H}$ are
linear, and if model errors are Gaussian (or zero), along with Gaussian
measurement errors with zero mean and known covariance, then the Kalman filter
provides the optimal state estimate ~\cite{kalman1960filter}. For nonlinear
problems, the extended Kalman filter (EKF)~\cite{ghil1981exkf} linearizes the
dynamics and observation operator so that the Kalman filter can be applied
locally. However, the cost of propagating the full state covariance matrix is
prohibitive for high-dimensional (in state space) problems. To address this
challenge, the EnKF~\cite{evensen1994sequential} approximates the covariance
using an ensemble of $N_e$ states
\begin{equation}
    Z_j=
    \begin{bmatrix}
        \bm{z}_{j}^{(1)},\cdots,\bm{z}_{j}^{(N_e)}
    \end{bmatrix}
    ,\quad j=1,\dots,J,
    \label{eq:ensForecastDef}
\end{equation}
where $Z_j\in\mathbb{R}^{N_z\times N_e}$. Similarly
we introduce
\begin{equation}
    \widehat{Z}_{j+1}=
    \begin{bmatrix}
        \widehat{\bm{z}}_{j+1}^{(1)},\cdots,\widehat{\bm{z}}_{j+1}^{(N_e)}
    \end{bmatrix}
    ,\quad j=0,\dots,J-1,\\
    \label{eq:ensForecastDef2}
\end{equation}
\begin{equation}
    \mathcal{H}\bigl(\widehat{Z}_{j+1}\bigr)=
    \begin{bmatrix}
        \mathcal{H}\bigl(\widehat{\bm{z}}_{j+1}^{(1)}\bigr),\cdots,
        \mathcal{H}\bigl(\widehat{\bm{z}}_{j+1}^{(N_e)}\bigr)
    \end{bmatrix}
    ,\quad j=0,\dots,J-1.
    \label{eq:ensForecastObsDef2}
\end{equation}
Using $\widehat{Z}_{j+1}$ and $\mathcal{H}\bigl(\widehat{Z}_{j+1}\bigr)$
the analysis operator for the EnKF may be defined and takes
    the form $\mathcal{A}:\mathbb{R}^{N_z\times N_e}\times\mathbb{R}^{N_d}\to
    \mathbb{R}^{N_z\times N_e}$. From it we obtain the analysis step
\begin{equation}
    Z_{j+1} = \mathcal{A}(\widehat{Z}_{j+1};\bm{d}_{j+1}),\quad j=1,\dots,J-1,
    \label{eq:enkfStdA}
\end{equation}
which updates the ensemble of states to obtain ${Z}_{j+1}$ from the predicted
states  $\widehat{Z}_{j+1}$ and the sensor data
$\bm{d}_{j+1}\in\mathbb{R}^{N_d}$. The standard EnKF cycle can then be
summarized as
\begin{equation}
    Z_{j+1} = \mathcal{A}\bigl(\mathcal{P}(Z_j);\bm{d}_{j+1}\bigr),
    \quad j=1,\dots,J-1,
    \label{eq:enkfStd}
\end{equation}
where we apply the prediction operator columnwise to $Z_j$.

We next describe how to realize the analysis operator $\mathcal{A}$
in~\eqref{eq:enkfStdA} using the EnKF. From the forecast ensemble
$\widehat{Z}_{j+1}$ we compute the ensemble mean
\begin{equation}
    \overline{\widehat{\bm{z}}}_{j+1}
    =
    \frac{1}{N_e}\sum_{n=1}^{N_e} \widehat{\bm{z}}_{j+1}^{(n)},
\end{equation}
and the ensemble perturbations
\begin{equation}
    \widehat{Z}'_{j+1}
    =
    \frac{1}{\sqrt{N_e-1}}
    \left[
        \widehat{\bm{z}}_{j+1}^{(1)} - \overline{\widehat{\bm{z}}}_{j+1},
        \ldots,
        \widehat{\bm{z}}_{j+1}^{(N_e)} - \overline{\widehat{\bm{z}}}_{j+1}
    \right].
\end{equation}
We also compute the predicted observations
\begin{equation}
    \widehat{Y}_{j+1}=
    \left[
        \widehat{\bm{y}}_{j+1}^{(1)},
        \ldots,
        \widehat{\bm{y}}_{j+1}^{(N_e)}
    \right]
    =
    \left[
        \mathcal{H}(\widehat{\bm{z}}_{j+1}^{(1)}),
        \ldots,
        \mathcal{H}(\widehat{\bm{z}}_{j+1}^{(N_e)})
    \right],
\end{equation}
where
\begin{equation}
    \bm{y}_j^{(n)} = \mathcal{H}(\bm{z}_j^{(n)}),\quad j=1,\dots,J,
    \label{eq:observations}
\end{equation}
so that $\widehat{Y}_{j+1}\in\mathbb{R}^{N_d\times N_e}$. From this we can
define the predicted observation mean
\begin{equation}
    \overline{\widehat{\bm{y}}}_{j+1}=
    \frac{1}{N_e}\sum_{n=1}^{N_e}\widehat{\bm{y}}_{j+1}^{(n)},
\end{equation}
and the associated perturbations
\begin{equation}
    \widehat{Y}'_{j+1}
    =
    \frac{1}{\sqrt{N_e-1}}
    \begin{bmatrix}
        \widehat{\bm{y}}_{j+1}^{(1)}-\overline{\widehat{\bm{y}}}_{j+1} &
        \dots &
        \widehat{\bm{y}}_{j+1}^{(N_e)}-\overline{\widehat{\bm{y}}}_{j+1}
    \end{bmatrix}.
\end{equation}
Let $R\in\mathbb{R}^{N_d\times N_d}$ denote the observation-error covariance,
and draw independent perturbations $\bm{\eta}_{j+1}^{(n)}\sim\mathcal{N}(0,R)$.
Then the forecast states are updated based on the predicted observations and the
observed data $\bm{d}_{j+1}\in\mathbb{R}^{N_d}$ as
\begin{equation}
    \bm{z}_{j+1}^{(n)}
    =
    \widehat{\bm{z}}_{j+1}^{(n)}
    +
    \widehat{Z}_{j+1}^{\prime}
    \widehat{Y}_{j+1}^{\prime T}
    \left(
        \widehat{Y}_{j+1}^{\prime}
        \widehat{Y}_{j+1}^{\prime T}
        +
        R
    \right)^{-1}
    \left(
        \bm{d}_{j+1}
        +
        \bm{\eta}_{j+1}^{(n)}
        -
        \widehat{\bm{y}}_{j+1}^{(n)}
    \right).
    \label{eq:EnKF_a}
\end{equation}
Stacking into a matrix we obtain
\begin{equation}
    Z_{j+1}=
    \begin{bmatrix}
        \bm{z}_{j+1}^{(1)},\cdots,\bm{z}_{j+1}^{(N_e)}
    \end{bmatrix}
    ,\quad j=1,\dots,J,
    \label{eq:ensAnalysisDef2}
\end{equation}
completing our definition of the analysis step
$\widehat{Z}_{j+1} \mapsto Z_{j+1}$: here, \eqref{eq:EnKF_a} is a concrete
realization of the analysis operator $\mathcal{A}$ in~\eqref{eq:enkfStdA}.

\begin{remark}\label{rmk:subspace}
Zhou \textit{et al.}~\cite{zhou2026neuralenkf} explain the failure of the
standard EnKF for problems with shocks in terms of pointwise violation of the
unimodal (near Gaussian) behavior that is implicit in its derivation.
An alternative perspective follows from function approximation. The EnKF
subspace property~\cite{evensen2003ensemble}, discussed for inverse problems
in~\cite{iglesias2013inverse}, implies that each analysis ensemble member
$z_{j+1}^{(m)}$ lies in the linear span of the forecast ensemble
$\widehat{Z}_{j+1}$. Consequently, applying the standard EnKF analysis step
directly in physical state space leads to approximation of shock profiles in the
linear span of other shock profiles. When shock locations differ across ensemble
members, these linear combinations distort sharp features; such features are
preserved only when they are aligned throughout the forecast ensemble.
\end{remark}

\subsection{Transformed EnKF} \label{ssec:tenkf}

The standard EnKF formulation in Subsection~\ref{ssec:senkf}, and in particular
the update formulae~\eqref{eq:EnKF_a} together with Remark~\ref{rmk:subspace},
make explicit that each analysis update lies in the linear span of the forecast
ensemble. This linear span is generally not well suited to approximating
functions with shocks and other discontinuities. The morphing 
EnKF~\cite{beezley2008registration,hu2026morphingenkf} and recent
neural-network-based
approaches~\cite{zhou2026neuralenkf,chandravamsi2026decoder} both address this
issue by applying the EnKF in a latent space. The morphing EnKF constructs a
latent representation consisting of a registration map and aligned state
representations, whereas the neural network-based methods employ learned latent
representations. This observation motivates a general framework for applying the
EnKF in feature-preserving latent spaces.

Let $\mathcal{T} : \mathbb{R}^{N_z}\to\mathbb{R}^{N_\phi}$ denote a
transformation from physical state space to a latent space of dimension
$N_\phi$ such that
\begin{equation}
\widehat{\bm{\phi}}_{j+1}^{(n)}
=
\mathcal{T}(\widehat{\bm{z}}_{j+1}^{(n)}),
\end{equation}
with $\widehat{\bm{\phi}}_{j+1}^{(n)}\in\mathbb{R}^{N_\phi}$. Assume that it
admits a reconstruction map
$\mathcal{T}^{-1}:\mathbb{R}^{N_\phi}\to\mathbb{R}^{N_z}$ with small
reconstruction error, $\|\mathcal{T}^{-1}(\mathcal{T}(\bm{z}))-\bm{z}\|_2^2$;
$\mathcal{T}^{-1}$ is thus an approximate inverse. The corresponding latent
space DA cycle is then
\begin{equation}
    Z_{j+1} = \mathcal{T}^{-1}\Bigl(
        \mathcal{A}\bigl(
        \mathcal{T}\bigl(
        \mathcal{P}(Z_j)
        \bigr);\bm{d}_{j+1}
        \bigr)
        \Bigr),\quad j=1,\dots,J-1.
    \label{eq:enkfTransformed}
\end{equation}
Here, $\mathcal{T}$ and $\mathcal{T}^{-1}$ are applied columnwise to the state
ensemble. Thus,~\eqref{eq:enkfTransformed} is the latent space analogue of the
standard EnKF cycle~\eqref{eq:enkfStd}.

Next, define the ensemble
\begin{equation}
    \widehat{\Phi}_{j+1}=
    \left[
        \widehat{\bm{\phi}}_{j+1}^{(1)},
        \ldots,
        \widehat{\bm{\phi}}_{j+1}^{(N_e)}
    \right],
\end{equation}
with mean
\begin{equation}
    \overline{\widehat{\bm{\phi}}}_{j+1}
    = \frac{1}{N_e}\sum_{n=1}^{N_e} \widehat{\bm{\phi}}_{j+1}^{(n)},
\end{equation}
and perturbations
\begin{equation}
    \widehat{\Phi}_{j+1}^{\prime}
    =
    \frac{1}{\sqrt{N_e-1}}
    \begin{bmatrix}
        \widehat{\bm{\phi}}_{j+1}^{(1)} - \overline{\widehat{\bm{\phi}}}_{j+1} &
        \cdots &
        \widehat{\bm{\phi}}_{j+1}^{(N_e)} - \overline{\widehat{\bm{\phi}}}_{j+1}
    \end{bmatrix}.
\end{equation}
Define
$\mathcal{H}_{\phi}:\mathbb{R}^{N_\phi}\to\mathbb{R}^{N_d}$ by
$\mathcal{H}_{\phi}=\mathcal{H} \circ \mathcal{T}^{-1}$.
Then we can compute the predicted observations from the transformed state using
\begin{equation}
    \widehat{\bm{y}}_{\phi,j+1}^{(n)}
    =
    \mathcal{H}_{\phi}(\widehat{\bm{\phi}}_{j+1}^{(n)}).
\end{equation}
This yields the ensemble-predicted observations
\begin{equation}
    \widehat{Y}_{\phi,j+1}=
    \left[
        \widehat{\bm{y}}_{\phi,j+1}^{(1)},
        \ldots,
        \widehat{\bm{y}}_{\phi,j+1}^{(N_e)}
    \right]
    =
    \left[
        \mathcal{H}_{\phi}(\widehat{\bm{\phi}}_{j+1}^{(1)}),
        \ldots,
        \mathcal{H}_{\phi}(\widehat{\bm{\phi}}_{j+1}^{(N_e)})
    \right],
\end{equation}
with mean
\begin{equation}
    \overline{\widehat{\bm{y}}}_{\phi,j+1}
    =\frac{1}{N_e}\sum_{n=1}^{N_e}\widehat{\bm{y}}_{\phi,j+1}^{(n)},
\end{equation}
and perturbations
\begin{equation}
    \widehat{Y}_{\phi,j+1}^{\prime}
    =
    \frac{1}{\sqrt{N_e-1}}
    \begin{bmatrix}
        \widehat{\bm{y}}_{\phi,j+1}^{(1)}-
        \overline{\widehat{\bm{y}}}_{\phi,j+1} &
        \dots &
        \widehat{\bm{y}}_{\phi,j+1}^{(N_e)}-
        \overline{\widehat{\bm{y}}}_{\phi,j+1}
    \end{bmatrix}.
\end{equation}
Then the analysis update in the latent space can be written
\begin{equation}
    \bm{\phi}_{j+1}^{(n)}
    =
    \widehat{\bm{\phi}}_{j+1}^{(n)}
    +
    \widehat{\Phi}_{j+1}^{\prime}
    \widehat{Y}_{\phi,j+1}^{\prime T}
    \left(
        \widehat{Y}_{\phi,j+1}^{\prime}
        \widehat{Y}_{\phi,j+1}^{\prime T}
        +
        R
    \right)^{-1}
    \left(
        \bm{d}_{j+1}
        +
        \bm{\eta}_{j+1}^{(n)}
        -
        \widehat{\bm{y}}_{\phi,j+1}^{(n)}
    \right).
    \label{eq:TransformedEnKF_a}
\end{equation}
We may stack these as in \eqref{eq:ensAnalysisDef2} to obtain $\Phi_{j+1}$,
completing our definition of the analysis step
$\widehat{\Phi}_{j+1} \mapsto \Phi_{j+1}$. We then recover the analysis state in
the original state space as $Z_{j+1}=\mathcal{T}^{-1}(\Phi_{j+1})$, thus
completing our definition of the transformed analysis step
$\widehat{Z}_{j+1} \overset{\mathcal{T}}{\mapsto} Z_{j+1}$. Note that the
sensor data $\bm{d}_{j+1}$ and observation-error covariance $R$ remain defined
in the original observation space.

\subsection{Constructing Latent Space Transformations}\label{ssec:tconstruct}

As outlined in Subsection~\ref{ssec:introLatentEnKF}, we seek a transformation
$\mathcal{T}$ with small reconstruction error, a feature-preserving
reconstruction map $\mathcal{T}^{-1}$, and a latent representation that varies
smoothly across ensemble members. The morphing
EnKF~\cite{beezley2008registration,hu2026morphingenkf} achieves these
properties through image registration: the latent variables comprise
aligned state representations and registration maps, which vary smoothly
across the ensemble, while the inverse transformation reconstructs sharp
features by warping the aligned state with the registration map. The
neural-network-based
approaches~\cite{zhou2026neuralenkf,chandravamsi2026decoder} instead
learn the latent transformation from data. Both neural-network-based approaches
include a reconstruction loss, while nonlinear activation functions promote
feature-preserving reconstructions. In addition, the neural
EnKF~\cite{zhou2026neuralenkf} encourages latent-space smoothness through
nearest-neighbor chain training, whereas the decoder-based 
approach~\cite{chandravamsi2026decoder} promotes it implicitly through the
joint learning of latent variables and a shared decoder.

In the morphing EnKF, the reconstruction map $\mathcal{T}^{-1}$ is obtained by
applying the inverse registration map to the aligned state representation,
whereas in the neural-network-based approaches it is specified by the
neural network architecture. In both cases, the corresponding encoding
map $\mathcal{T}$ is defined implicitly, either by solving a registration
problem or through neural network training. Following this perspective,
throughout this paper, we specify $\mathcal{T}^{-1}$ explicitly and define
$\mathcal{T}$ implicitly through an optimization problem.

\subsection{Transformed EnKF for a Hyperbolic Tangent Profile}
\label{ssec:parametrizedHeaviside}

Consider the parameterized hyperbolic tangent
\begin{equation}
    H_\delta(x;c_{L},c_{R},x_{s})=
    \frac{c_L+c_R}{2}-\frac{c_L-c_R}{2}
    \tanh\left({\frac{x-x_s}{\delta}}\right).
    \label{eq:heaviside}
\end{equation}
We define the reconstruction
$\mathcal{T}^{-1}: \mathbb{R}^{N_\phi} \to C(\Omega)$ by
\begin{subequations}
\begin{equation}
    \mathcal{T}^{-1}(\bm{\phi})(x)
    =
    H_{\delta}(x;{c}_L,{c}_R,{x}_s),
\end{equation}
with transformed coordinates
\begin{equation}
    \bm{\phi}=
    ({c}_L, {c}_R, {x}_s, \delta)\in\mathcal{X}
\end{equation}
and latent space
\begin{equation}
    \mathcal{X}\equiv \mathbb{R}\times\mathbb{R}
    \times\Omega\times\mathbb{R}_{>0}.
    \label{eq:HwLatent}
\end{equation}
\label{eq:HwReconstruction}
\end{subequations}
We then define $\mathcal{T}: C(\Omega) \to \mathbb{R}^{N_\phi}$ 
through the minimization problem
\begin{subequations}
\begin{equation}
  \bm{\phi}_*=\argmin_{\bm{\phi}\in\mathcal{X}}J(\bm{\phi};c_L,c_R,x_s,
  \delta),
\end{equation}
with objective function
\begin{equation}
  J(\bm{\phi};c_L,c_R,x_s,\delta)
  =
  \int_{\Omega}
  \left|
  \mathcal{T}^{-1}(\bm{\phi})(x)-H_\delta(x;c_L,c_R,x_s)
  \right|^2 dx.
\end{equation}
so that
\begin{equation}
    \mathcal{T}\big(H_\delta(x;c_L,c_R,x_s)\big)=(c_L,c_R,x_s,\delta).
\end{equation}
\label{eq:HwOptim}
\end{subequations}
No regularization is needed to promote latent space smoothness because, in
this special case, the reconstruction map exactly matches the parameterization
used to generate the data, whose latent variables are Gaussian.

We now demonstrate the transformed EnKF for~\eqref{eq:heaviside} with
reconstruction function~\eqref{eq:HwReconstruction}. We set $\Omega\equiv[0,1]$
with grid $x_i=i\Delta x$ for $i=0,\dots,N_z-1$, where $\Delta x = 1/(N_z-1)$,
so that $\bm{x}\in\mathbb{R}^{N_z}$. In the numerical experiments presented
below we take $N_z=400$ and $\delta=4\Delta x$, with the same width for both the
reference truth and all ensemble members. We define a reference truth profile
$\bm{z}^\dagger=H_\delta(\bm{x};c^\dagger_L,c^\dagger_R,x^\dagger_s)$, with
$c^\dagger_L=2$, $c_R^\dagger=1$, and $x^\dagger_s=0.55$, which we will observe
noisily at four points evenly distributed between $x=0.2$ and $x=0.8$. Let
$\bm{x}_{\mathrm{obs}}\in\mathbb{R}^{N_d}$ denote these points, with
\begin{equation}
    x_{\mathrm{obs},i}=0.2+i0.2,\quad i=0,\dots,3.
\end{equation}
The corresponding data vector $\bm{d}\in\mathbb{R}^{N_d}$ has entries
\begin{equation}
    d_i = H_\delta(x_{\mathrm{obs},i},
    c^\dagger_L,c^\dagger_R,x^\dagger_s)+
    \eta_i,\quad i=0,\dots,N_d-1,
\end{equation}
where $\bm{\eta}\sim\mathcal{N}(0,R)$ for the observation-error covariance $R$,
whose diagonal entries are given by $R_{ii}=\max\{0.1 d_i,0.05\}$. This
choice imposes a relative observation error of 10\% and an absolute error of
0.05.

\begin{remark}
    In Subsection~\ref{ssec:senkf}, we introduce the state vector
    $\bm{z}_{j+1}\in\mathbb{R}^{N_z}$. For one-dimensional problems, $N_z$ is
    equal to the number of grid points in the $x$ direction. We use the notation
    $N_z$ instead of $N_x$ for consistency with Subsection~\ref{ssec:senkf}.
\end{remark}

We construct an ensemble of $N_e=30$ members by sampling, for
$n=1,\dots,N_e$,
\begin{equation}
\begin{aligned}
c_{L}^{(n)}&\sim\mathcal{N}(2,0.2^2),&
c_{R}^{(n)}&\sim\mathcal{N}(1,0.1^2),&
x_{s}^{(n)}&\sim\mathcal{N}(0.5,0.05^2),
\end{aligned}
\end{equation}
and setting
$\bm{z}^{(n)}=H_\delta(\bm{x};c_{L}^{(n)},c_{R}^{(n)},x_{s}^{(n)})$. Thus, the
ensemble-mean shock location,
\[
\overline{x}_s=\frac{1}{N_e}\sum_{n=1}^{N_e}x_s^{(n)}=0.5,
\]
differs from the reference truth value $x_s^\dagger=0.55$.

\begin{figure}
    \centering
    \begin{subfigure}{0.35\linewidth}
        \centering
        \includegraphics[width=\linewidth]{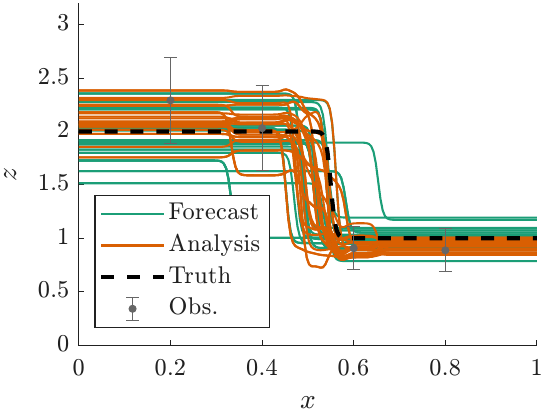}
        \caption{Standard EnKF}
        \label{fig:lsDemoH3}
    \end{subfigure}
    \hspace{0.04\linewidth}
    \begin{subfigure}{0.35\linewidth}
        \centering
        \includegraphics[width=\linewidth]{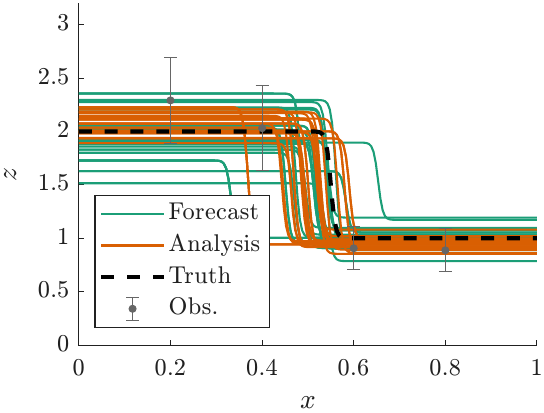}
        \caption{Transformed EnKF}
        \label{fig:lsDemoH2}
    \end{subfigure}
    \caption{Forecast and analysis ensembles for the parameterized hyperbolic
    tangent profile~\eqref{eq:heaviside}. The transformed EnKF is
    feature-preserving, while the standard EnKF introduces spurious
    oscillations.}
    \label{fig:lsDemoH}
\end{figure}

Figure~\ref{fig:lsDemoH3} shows that, although the forecast ensemble consists
entirely of hyperbolic tangent profiles, the analysis ensemble produced by
the standard EnKF~\eqref{eq:EnKF_a} does not. The linear EnKF update in the
physical state space introduces spurious oscillations near the discontinuity
(see Remark~\ref{rmk:subspace}). In contrast, Figure~\ref{fig:lsDemoH2} shows
that performing the analysis step with the transformed
EnKF~\eqref{eq:TransformedEnKF_a} in the latent variables defined
by~\eqref{eq:HwOptim} preserves the hyperbolic tangent structure. This is
because the reconstruction map~\eqref{eq:HwReconstruction} enforces hyperbolic
tangent profiles by construction.

\section{Level Set Representations for 1D Scalar Fields}\label{sec:ls1d}

In this section, we generalize the transformed EnKF approach of
Subsection~\ref{ssec:parametrizedHeaviside} by allowing the left and right
states to vary with $x$, rather than remain constant, to better represent the
spatially varying functions encountered in compressible flows. We formulate a
regularized optimization problem to identify the level set representation, such
that uncertainty in the shock location is decoupled from uncertainty in the
solution on either side of the shock. This decoupling yields a representation
that varies smoothly across ensembles of similar solutions, and hence identifies
a latent space that is favorable for the analysis step in the DA cycle. In 
Subsection~\ref{ssec:ls1d-1}, we consider problems with a single discontinuity
and define a regularized optimization problem that ensures the desired
properties of the level set representation. In Subsection~\ref{ssec:ls1d-K} we
extend the methodology to multiple discontinuities. The key ideas, namely the
level set representation and the optimization problem to determine it, are
contained in the main body of the text; details of our specific implementation
of the methodology are presented in~\ref{app:ls1dDetails}.

\subsection{1D Level Set Representation for a Scalar Field
 with a Single Discontinuity}\label{ssec:ls1d-1}

For a single discontinuity in 1D, we propose the reconstruction map
\begin{subequations}
\begin{equation}\label{eq:tanh}
  \mathcal{T}^{-1}(\bm{\phi})(x)
  =\frac{{f}_{L}(x)+{f}_{R}(x)}{2}
  -\frac{{f}_{L}(x)-{f}_{R}(x)}{2}
  \tanh\left({\frac{x-{x}_s}{{\delta}}}\right),
\end{equation}
where
\begin{equation}
  \bm{\phi}=({f}_L,{f}_R,{x}_s,\delta),
  \qquad
  \bm{\phi}\in\mathcal{X}
\end{equation}
for
\begin{equation}
  \mathcal{X}\equiv
  C^1(\Omega)\times C^1(\Omega)\times\Omega\times\mathbb{R}_{>0}.
\end{equation}
\label{eq:tanh_ls}
\end{subequations}
Here, $C^1(\Omega)$ is the space of functions on $\Omega$ with continuous first
derivatives, ${f}_{L}$ and ${f}_{R}$ are smooth, globally defined,
state extensions to the left and right of the discontinuity, respectively,
${x}_s$ is an estimate of the discontinuity location, and ${\delta}$ is an
estimate of its width.

\begin{remark}
As discussed in Subsection~\ref{ssec:introLatentEnKF}, we seek a
feature-preserving latent representation, so that perturbations in latent
variables preserve sharp features in physical space. In the present setting,
this property is achieved through the nonlinear reconstruction
map~\eqref{eq:tanh}, which introduces nonlinearity through the hyperbolic
tangent function. Similarly, neural network-based approaches obtain nonlinear
reconstruction maps through their activation functions, while the morphing EnKF
does so through its registration map. This nonlinearity is important because
the standard EnKF analysis update is confined to the linear span of the forecast
ensemble; as discussed in Remark~\ref{rmk:subspace}, linear combinations of states
with misaligned discontinuities generally do not preserve sharp features. Thus,
for ensembles with uncertain discontinuity locations, feature-preserving latent
representations generally require a nonlinear reconstruction map.
\end{remark}

\begin{remark}\label{rmk:levelSet1D}
We refer to~\eqref{eq:tanh} as a level set representation because the
discontinuity is encoded by the location $x_s$, which is the zero level set of
the signed distance function
\begin{equation}\label{eq:levelSet1D}
  \varphi(x)=x-x_s.
\end{equation}
We do not explicitly use this level set function in 1D, since the discontinuity
location can be parameterized directly by $x_s$. In 2D, however, the
discontinuity is a curve, so we parameterize it by a level set function rather
than by a sequence of points; see Section~\ref{sec:ls2dEuler}.
\end{remark}

\subsubsection{Regularized Optimization Problem For
The Level Set Representation}\label{sssec:rop}

We seek $x_s$, $f_L$, and $f_R$ such that the reconstruction
map~\eqref{eq:tanh_ls} accurately approximates the function $f$ while the
associated level set representations vary smoothly across an ensemble of similar
functions. In particular, we seek to encode all information about the shock
location in $x_s$, rather than in $f_L$ and $f_R$. To achieve this, we solve the
regularized minimization problem
\begin{subequations}
\begin{equation}
  \bm{\phi}_*=\argmin_{\bm{\phi}\in\mathcal{X}}J(\bm{\phi};f),
\end{equation}
with objective function
\begin{equation}
  \begin{aligned}
  J(\bm{\phi})
  &=
  \int_{\Omega}
  \left|
  \mathcal{T}^{-1}(\bm{\phi})(x)-f(x)
  \right|^2 dx\\
  &+
  \lambda_1
  \int_{\Omega}
  \left(
  |{f}_L'(x)|^2
  +
  |{f}_R'(x)|^2
  \right) dx\\
  &+\lambda_b(|f_L(a)-f(a)|^2+|f_R(b)-f(b)|^2).
  \end{aligned}
\end{equation}
\label{eq:1dLSObjCont}
\end{subequations}
Here, $\lambda_1$ and $\lambda_b$ are regularization parameters, where
$\lambda_1$ promotes smoothness while $\lambda_b$ enforces boundary conditions.

We solve~\eqref{eq:1dLSObjCont} using a bound-constrained trust-region
reflective nonlinear least-squares method, with the Jacobian approximated using
two-point finite differences. However, directly solving this problem from an
arbitrary initialization can converge slowly. We therefore introduce several
modifications to reduce computational cost and improve convergence, as detailed
in \ref{app:ls1dDetails}. We briefly summarize the main ideas here.

Although~\eqref{eq:1dLSObjCont} is formulated globally in $x$, we show in
\ref{sapp:ls1dDetailsLocalProblemCont} how to formulate a local optimization
problem over a neighborhood of the discontinuity. This reduces the number of
unknowns and residuals, yielding a smaller Jacobian and decreasing the
computational cost. We first estimate $x_s$ using the gradient of $f$ and
identify a neighborhood about $x_s$ bounded by local minima of the gradient
magnitude. We also construct smooth initial guesses for $f_L$ and $f_R$ that
yield a small reconstruction error. These choices provide an informed
initialization, and the optimization then refines the representation within this
neighborhood. We impose the bounds $\delta\in[0.1\Delta x,100\Delta x]$ and
$x_s\in\Omega$, while leaving $f_L$ and $f_R$ unbounded.

\begin{remark}\label{rmk:regularizationFail}
Note that, without regularization, one can obtain zero reconstruction error by
taking ${f}_{L}(x)=f(x)$ and ${f}_{R}(x)=f(x)$, with any $\delta>0$ and
${x}_s\in\Omega$. However, such a solution encodes the shock location in $f_L$
and $f_R$, rather than solely in $x_s$. Consequently, $f_L$ and $f_R$ will not
vary smoothly across an ensemble of similar functions, and the resulting level
set representation fails to satisfy the desired criteria.
\end{remark}

\begin{remark}
The parameters $\lambda_1$ and $\lambda_b$ balance reconstruction accuracy,
smoothness of the state extensions, and agreement with the boundary values. As
discussed in Remark~\ref{rmk:regularizationFail}, insufficient regularization
allows the state extensions to encode information about the shock location,
which we wish to avoid. In contrast, excessively large values of $\lambda_1$
yield smooth but inaccurate state extensions. We therefore seek the smallest
$\lambda_1$ that suppresses shock-like gradients in $f_L$ and $f_R$ while
yielding accurate reconstruction. The parameter $\lambda_b$ controls agreement
with the boundary values, which are imposed through a penalty rather than as
hard constraints. It can be increased until the boundary mismatch falls below a
user-prescribed tolerance. The regularization parameters are problem specific
and should be tuned on representative ensemble members, assessing both
reconstruction accuracy and smoothness of the state extensions across ensemble
members. Representative values for 1D compressible flows are presented in
Table~\ref{tab:1d-euler-parameters} of Subsection~\ref{ssec:ls1d:description}.
\end{remark}

\begin{remark}\label{rmk:jacobian}
For simplicity, we approximate the entire Jacobian using finite differences.
Localizing the optimization problem substantially reduces the cost of this
approximation. Further efficiencies could be achieved by exploiting the
structure of the level set representation. In particular, the derivative of
the hyperbolic tangent is available analytically, while the Jacobians of the
smooth functions on either side of the discontinuity could be computed using
automatic differentiation. Alternatively, the smooth functions could be
represented using a low-dimensional basis, resulting in lower-dimensional
Jacobians that can be evaluated analytically.
\end{remark}

\subsubsection{Demonstration of the The Level Set Representation}

\begin{figure}
    \centering
    \includegraphics[width=0.6\linewidth]{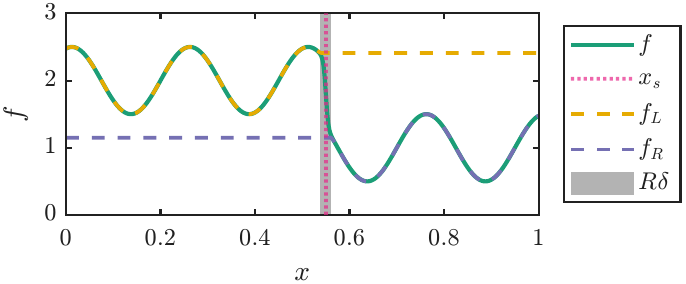}
    \caption{Identifying the level set representation of~\eqref{eq:lsDemoSin} by
    solving~\eqref{eq:1dLSObjCont}. Implementation details are presented in
    \ref{app:ls1dDetails}. We visualize the width by shading 
    $x\in[x_s-R\delta/2,x_s+R\delta/2]$.}
    \label{fig:lsDemoSin_schematic}
\end{figure}

Here, we demonstrate the level set representation~\eqref{eq:tanh} for
\begin{equation}
    f(x)
    =
    \frac{3}{2}
    +
    \frac{1}{2}\sin(8\pi(x-0.2))
    -
    \frac{1}{2}
    \tanh\left(\frac{x-x^\dagger_s}{\delta^\dagger}\right)
    \label{eq:lsDemoSin}
\end{equation}
defined over $\Omega\equiv[0,1]$, with discontinuity location
$x_s^\dagger=0.5$, and width $\delta^\dagger=0.02$.
Figure~\ref{fig:lsDemoSin_schematic} depicts~\eqref{eq:lsDemoSin} along with its
level set representation, showing ${f}_L$, ${f}_R$, and ${x}_s$. We
visualize the width by shading the region between the two points where
$\tanh((x-x_s)/\delta)$ achieves 1\% and 99\% of its maximum value,
denoted. This region has width $R\delta$, for $R\approx4.60$, and the
corresponding interval is $[x_s-R\delta/2,x_s+R\delta/2]$.
Figure~\ref{fig:lsDemoSin_schematic} highlights an advantage of the level set
EnKF: its latent representation is physically interpretable. The resulting
interpretability facilitates understanding the action of the filter and enables
the incorporation of problem-specific inductive biases.

\subsection{1D Level Set Representation for a Scalar Field with
Multiple Discontinuities} \label{ssec:ls1d-K}

This subsection modifies the 1D level set procedure presented in
Subsection~\ref{ssec:ls1d-1} so that it can be applied to a scalar field
$f(x)$ with multiple discontinuities. Throughout this subsection, we will
assume that the number of discontinuities, $K$, is known \textit{a priori}.

For $K$ discontinuities in one space dimension, our reconstruction map becomes
\begin{subequations}
\begin{equation}
  \mathcal{T}^{-1}(\bm{\phi})(x)
  =
  \sum_{k=0}^{K}
  \alpha_k(x){f}_k(x),
\end{equation}
where
\begin{equation}
  \begin{aligned}
    \alpha_0(x)
    &=
    \prod_{j=1}^{K} \left(1-H_j(x)\right),\\
    \alpha_k(x)
    &=
    \left(
    \prod_{j=1}^{k} H_j(x)
    \right)
    \left(
    \prod_{j=k+1}^{K} \left(1-H_j(x)\right)
    \right),
    \quad k=1,\dots,K-1,\\
    \alpha_K(x)
    &=
    \prod_{j=1}^{K} H_j(x),
  \end{aligned}
\end{equation}
for
\begin{equation}
  H_j(x)
  =
  \frac{1}{2}
  \left[
  1+
  \tanh\left(
  \frac{x-{x}_{s,j}}{{\delta}_j}
  \right)
  \right],
  \quad j=1,\dots,K.
\end{equation}
The latent variables are then
\begin{equation}
  \bm{\phi}
  =
  \left(
  {f}_0,\ldots,{f}_K,
  {x}_{s,1},\ldots,{x}_{s,K},
  {\delta}_1,\ldots,{\delta}_K
  \right),
\end{equation}
where
\begin{equation}
  \bm{\phi}\in
  \mathcal{X}_K,\quad
  \mathcal{X}_K\equiv
  \left(C^1(\Omega)\right)^{K+1}
  \times \Omega_<^K
  \times \mathbb{R}_{>0}^K,
\end{equation}
with
\begin{equation}
  \Omega_<^K
  =
  \{({x}_{s,1},\ldots,{x}_{s,K})\in\Omega^K:
  {x}_{s,1}<\cdots<{x}_{s,K}\}.
\end{equation}
\end{subequations}

In principle, we wish to solve the regularized minimization problem
\begin{subequations}
\begin{equation}
  \bm{\phi}_*=\argmin_{
  \bm{\phi}
  \in
  \mathcal{X}_K
  }
  J(\bm{\phi};f),
\end{equation}
with objective function
\begin{equation}
  \begin{aligned}
  J(\bm{\phi})
  &=
  \int_{\Omega}
  \left|
  \mathcal{T}^{-1}(\bm{\phi})(x)-f(x)
  \right|^2 dx\\
  &+
  \lambda_1
  \sum_{k=0}^{K}
  \int_{\Omega}
  \left(
  |{f}_k'(x)|^2
  \right) dx\\
  &+\lambda_b(|f_0(a)-f(a)|^2+|f_K(b)-f(b)|^2).
  \end{aligned}
\end{equation}
\label{eq:1dLSObjCont-K}
\end{subequations}
Although we could attempt to directly solve~\eqref{eq:1dLSObjCont-K} using a
nonlinear least-squares solver, it is more effective to reformulate it as
multiple local optimization problems. To do so, we first estimate the centers of
the $K$ discontinuities using the local maxima of the gradient of $f$. We then
define a neighborhood around each discontinuity, bounded by local minima of the
gradient of $f$. This decomposition allows us to solve $K$ local regularized
minimization problems independently. As in the case of a single discontinuity
described in Subsubsection~\ref{sssec:rop}, the optimization need only determine
a small local correction to $\bm{\phi}$ near each of the $K$ discontinuities,
rather than optimize from an arbitrary initial guess globally in $x$. We provide
a detailed description of this procedure in \ref{sapp:1dLSDetails-K}.
\section{Level Set EnKF for 1D Compressible Flows}
\label{sec:enkf1dEuler}
\label{sec:ls1dEuler}

This section is concerned with applying the level set EnKF to 1D compressible
Euler problems. Subsection~\ref{ssec:ls1d:description} describes the
demonstration problems, Subsection~\ref{ssec:setup1d} describes the setup
of our filtering problem and how we evaluate EnKF performance.
Subsection~\ref{ssec:ls1dCompFlow} extends the level set representation
procedure of Section~\ref{sec:ls1d} to solutions to the 1D compressible Euler
equations, while Subsection~\ref{ssec:ls1dEnKF} describes how to construct level
set representations for an ensemble of flow states, and then how to apply the
EnKF analysis step~\eqref{eq:TransformedEnKF_a} to the resulting level set
representations. Finally, Subsection~\ref{ssec:1dResults} demonstrates the level
set EnKF for the problems outlined in Subsection~\ref{ssec:ls1d:description}.

\subsection{Description of Demonstration Problems}\label{ssec:ls1d:description}

We apply the level set EnKF to three solutions of the 1D compressible Euler
equations:
Toro's shock tube problem~\cite{toro2009compressible},
the Shu-Osher shock-entropy problem~\cite{shu1989eno}, and
Sod's shock tube problem~\cite{sod1978shockTube}. These problems were
previously studied by Subrahmanya and Sandu~\cite{subrahmanya2025etpfdtw} in
their development of the ETPF-DTW approach. We use the same problems because,
together, they contain the canonical flow features present in compressible
flows: shocks, contact surfaces, and rarefaction waves.

The initial conditions for our demonstration problems take the form
\begin{equation}\label{eq:riemannProblem1d}
    \bm{w}(x) =
    \begin{cases}
        \bm{w}_L(x), & x \leq x_d, \\
        \bm{w}_R(x), & x > x_d,
    \end{cases},
    \quad x\in\Omega.
\end{equation}
for $\Omega\equiv[0,1]$. Here, $\bm{w} = (\rho,u,p)$ denotes the vector of
primitive variables and $x_d$ denotes the location of a diaphragm separating the
left state $\bm{w}_L(x)$ from the right state $\bm{w}_R(x)$. Thus, the initial
condition $\bm{w}(x)$ is discontinuous at $x = x_d$. At time $t = 0$, the
diaphragm is removed, allowing the two states to interact.

We discretize the spatial domain $\Omega$ uniformly using $N_x$ grid points:
\begin{equation}\label{eq:gridX}
    x_i = i\Delta x,
    \qquad i = 0,\ldots,N_x-1,
    \qquad \Delta x = \frac{1}{N_x-1}.
\end{equation}
The resulting vector of grid-point coordinates is
$\bm{x} \in \mathbb{R}^{N_x}$. We specify problem-specific final times, $T$, and
choose time steps satisfying the CFL condition. For each calculation, the values
of $N_x$ and $T$ are listed in Table~\ref{tab:1d-euler-parameters}.

Subsubsections~\ref{sssec:ls1dToro}, \ref{sssec:ls1dShuOsher},
and~\ref{sssec:ls1dSod}, describe the initial conditions and
resulting solutions for Toro's shock tube problem, the Shu-Osher shock-entropy
problem, and Sod's shock tube problem, respectively. The variables used to
define the initial conditions are presented in
Table~\ref{tab:1d-euler-reference-parameters}.

\begin{table}
    \centering
    \small
    \renewcommand{\arraystretch}{1.25}
    \begin{tabular}{@{}l|ccc|c|cc@{}}
        \hline
        & $T$ & $t_{\mathrm{DA},1}$ & $\Delta t_{\mathrm{DA}}$ &
        $N_x$ & $\lambda_1$ & $\lambda_b$ \\
        \hline
        Toro      & $0.399$ & $0.007$ & $0.0035$ & $400$ & $10^{2}$ & $10^{2}$ \\
        Shu-Osher & $0.25$  & $0.025$ & $0.0125$ & $800$ & $10^{-1}$ & $10^{2}$ \\
        Sod       & $0.2$   & $0.06$  & $0.01$   & $400$ & $10^2$ & $10^2$ \\
        \hline
    \end{tabular}
    \caption{Final time $T$, time of first DA cycle $t_{\mathrm{DA},1}$,
    spacing of subsequent DA cycles $\Delta t_{\mathrm{DA}}$, spatial resolution
    $N_x$, and level set regularization parameters $\lambda_1$ and $\lambda_b$
    for the 1D Euler experiments.}
    \label{tab:1d-euler-parameters}
\end{table}

\subsubsection{Toro's Shock Tube Problem}\label{sssec:ls1dToro}

Toro~\cite{toro2009compressible} introduces a shock tube problem with initial
left and right states that are spatially constant. The resulting solution
comprises two shocks and a contact surface, all traveling to the right.

\subsubsection{Shu-Osher Shock-Entropy Problem}\label{sssec:ls1dShuOsher}

The solution to Toro's shock tube problem is nearly piecewise constant, so we
apply our method to the Shu-Osher shock-entropy problem~\cite{shu1989eno} to
highlight that it can be applied to more complicated functions. All fields 
of the initial left and right states are spatially constant except for the
density field to the right of the diaphragm, which varies sinusoidally as
\begin{equation}\label{eq:shuOsherRhoR}
    \rho_R(x) = \widetilde{\rho}_R
        + 0.2\sin\left(10\pi(x-x_d)\right).
\end{equation}
The solution comprises a right-going shock interacting with an entropy wave.

\subsubsection{Sod's Shock Tube Problem}\label{sssec:ls1dSod}

Sod's shock tube problem~\cite{sod1978shockTube} serves as our final
demonstration problem. All fields of the initial left and right states are
spatially constant, and the solution comprises a left-going rarefaction wave, a
right-going contact surface, and a right-going shock.

\subsection{Numerical Setup and Performance Metrics}\label{ssec:setup1d}

Subsubsection~\ref{sssec:setup1dFilter} describes the filtering problem
and observation setting for the 1D Euler experiments, while
Subsubsection~\ref{sssec:setup1dMetrics} describes the performance
metrics used to assess level set EnKF.

\subsubsection{Filtering Problem and Observation Setting}\label{sssec:setup1dFilter}

Uncertainty in the initial condition leads to uncertainty in our state estimate,
which we reduce by solving the filtering problem using the transformed
EnKF~\eqref{eq:TransformedEnKF_a}, with the level set representation as
our latent variables. We assume that the parameters specifying the true initial
condition~\eqref{eq:riemannProblem1d} are known to problem-specific tolerances.
Let $X$ denote an uncertain parameter. For each such parameter, we draw each
ensemble member independently according to
\begin{equation}\label{eq:randomVar}
    X^{(n)}\sim\mathcal{N}(\mu_X,\sigma_X^2),\quad n=1,\dots,N_e,
\end{equation}
where $\mu_X$ and $\sigma_X$ denote the mean and standard deviation,
respectively. Table~\ref{tab:1d-euler-reference-parameters} reports these values
along with the reference truth. Note that the ensemble means match the truth
values for the left and right states, while the diaphragm-location means are
intentionally offset from the truth to increase the uncertainty in the
discontinuity locations.

\begin{table}
    \centering
    \small
    \renewcommand{\arraystretch}{1.25}
    \setlength{\tabcolsep}{3pt}
    \begin{tabular}{@{}ll|ccc|ccc|c@{}}
        \hline
        Problem & Statistic & $\rho_L$ & $u_L$ & $p_L$
        & $\rho_R$ & $u_R$ & $p_R$ & $x_d$ \\
        \hline
        Toro & Truth
        & $5.99924$ & $19.5975$  & $460.894$
        & $5.99242$ & $-6.19633$ & $46.0950$
        & $0.41$ \\
        & $\mu$
        & $\rho_L^\dagger$ & $u_L^\dagger$  & $p_L^\dagger$
        & $\rho_R^\dagger$ & $u_R^\dagger$ & $p_R^\dagger$
        & $0.5$ \\
        & $\sigma$
        & $0.2$ & $0.0$ & $46.0894$
        & $0.1$ & $0.0$ & $4.6095$
        & $0.1$ \\
        \hline
        Shu-Osher & Truth
        & $3.857143$ & $2.629369$ & $10.33333$
        & $1.0$      & $0.0$      & $1.0$
        & $0.05$ \\
        & $\mu$
        & $\rho_L^\dagger$ & $u_L^\dagger$ & $p_L^\dagger$
        & $\rho_R^\dagger$ & $u_R^\dagger$ & $p_R^\dagger$
        & $0.1$ \\
        & $\sigma$
        & $0.4$ & $0.2$ & $1.03333$
        & $0.1$ & $0.0$ & $0.1$
        & $0.04$ \\
        \hline
        Sod & Truth
        & $1.0$   & $0.0$ & $1.0$
        & $0.125$ & $0.0$ & $0.1$
        & $0.59$ \\
        & $\mu$
        & $\rho_L^\dagger$ & $u_L^\dagger$ & $p_L^\dagger$
        & $\rho_R^\dagger$ & $u_R^\dagger$ & $p_R^\dagger$
        & $0.5$ \\
        & $\sigma$
        & $0.05$  & $0.0$ & $0.05$
        & $0.006$ & $0.0$ & $0.005$
        & $0.1$ \\
        \hline
    \end{tabular}
    \caption{Truth parameters, as well as the means and standard deviations used
    to construct the initial ensembles for the 1D Euler experiments. For the
    Shu-Osher shock-entropy problem, $\rho_R$ denotes $\tilde{\rho}_R^\dagger$
    in~\eqref{eq:shuOsherRhoR}.}
    \label{tab:1d-euler-reference-parameters}
\end{table}

We place pressure sensors at
\begin{equation}
    x_{\mathrm{obs},i}=0.2+i0.2,\quad i=0,\dots,3,
\end{equation}
so that $N_d=4$. The data vector $\bm{d}\in\mathbb{R}^{N_d}$ has entries
\begin{equation}
    d_i = p^\dagger(x_{\mathrm{obs},i},t;
    \bm{w}_L^\dagger,\bm{w}_R^\dagger,x_d^\dagger)+\eta_i,
    \quad i=0,\dots,N_d-1,
\end{equation}
where $\bm{\eta}\sim\mathcal{N}(0,R)$ and
$R_{ii}=\max\{0.1p^\dagger(x_{\mathrm{obs},i},t;
\bm{w}_L^\dagger,\bm{w}_R^\dagger,x_d^\dagger),0.05\}$, corresponding to 10\%
relative observation error with a minimum absolute error of 0.05. The reference
pressure field $p^\dagger$ is obtained by solving the Euler equations
numerically on a grid of size $N_x^\dagger=4000$ and interpolating the solution
to the observation point $x_{\mathrm{obs},i}$.

We solve the filtering problem using ensembles of size $N_e=50$, with the DA
schedules listed in Table~\ref{tab:1d-euler-parameters}, where
$t_{\mathrm{DA},1}$ denotes the time of the first DA cycle and
$\Delta t_{\mathrm{DA}}$ denotes the spacing between subsequent DA cycles.
Table~\ref{tab:1d-euler-parameters} also lists the regularization parameters
$\lambda_1$ and $\lambda_b$ used in~\eqref{eq:1dLSObjCont} to construct the
level set representation.

\subsubsection{Performance Metrics}\label{sssec:setup1dMetrics}

For an arbitrary field $\bm{f}_j\in\mathbb{R}^{N_x}$ at time index $j$, we compute the root mean square error (RMSE)
\begin{equation}
    \mathrm{RMSE}(\bm{f}_j)=\frac{1}{\sqrt{N_x}}\|\overline{\bm{f}}_{j}-\bm{f}^{\dagger}_{j}\|_2,
\end{equation}
which measures the error between the ensemble mean and reference truth, while we compute the
ensemble spread as
\begin{equation}
\mathrm{Spread}(\bm{f}_j)=
\sqrt{\frac{1}{N_x}\mathrm{trace}(P\left(\bm{f}_j)\right)},\quad
P(\bm{f}_j)=\frac{1}{N_e-1}\sum_{n=1}^{N_e}
(\bm{f}^{(n)}_j-\overline{\bm{f}_j})(\bm{f}_j^{(n)}-\overline{\bm{f}_j})^T,
\end{equation}
which measures the uncertainty represented by the ensemble. The EnKF approximates the
uncertainty in the state estimate using the sample covariance of an ensemble. If
the EnKF is well calibrated, then the ensemble spread should provide a
reasonable estimate of the actual estimation error. Consequently, the ensemble
spread and the RMSE should be of similar magnitude, and we assess performance by
plotting both RMSE and ensemble spread as functions of DA cycle. We additionally
define the \textit{farthest ensemble member} to be the member with the largest
Euclidean distance from the ensemble mean.

\subsection{Level Set Representations for 1D Compressible Flows}
\label{ssec:ls1dCompFlow}

For compressible flows, we apply the level set representation developed in
Section~\ref{sec:ls1d} field-by-field to the primitive variables
$\bm{w}$, computing the representation for each field independently. The
resulting latent space representation for field $\bm{w}_i\in\mathbb{R}^{N_x}$
with $K_i$ discontinuities is
\begin{subequations}\label{eq:latent}
\begin{equation}
    \bm{\phi}_i =
    \begin{bmatrix}
        \bm{w}_{i,1} & \dots & \bm{w}_{i,K_i+1} &
        \bm{\varphi}_{i,1} & \dots & \bm{\varphi}_{i,K_i} &
        \delta_{i,1} & \dots & \delta_{i,K}
    \end{bmatrix}^T,
\end{equation}
where $\bm{w}_{i,k}\in\mathbb{R}^{N_x}$ for $k=1,\dots,K_i+1$,
$\bm{\varphi}_{i,k}\in\mathbb{R}^{N_x}$ for $k=1,\dots,K_i$, and
$\delta_{i,k}\in\mathbb{R}_{>0}$ for $k=1,\dots,K_i$. We then concatenate
$\bm{\phi}_i\in\mathbb{R}^{(2K_i+1)\times N_x+K_i}$ for $i=1,2,3$ to obtain
\begin{equation}
    \bm{\phi} =
    \begin{bmatrix}
        \bm{\phi}_1 & \bm{\phi}_2 & \bm{\phi}_3
    \end{bmatrix}^T,
\end{equation}
\end{subequations}
the level set representation of $\bm{w}$. Implementation details are presented
in \ref{sapp:1d-comp}.

\begin{figure}
    \centering
    \begin{subfigure}[t]{0.32\linewidth}
        \centering
        \includegraphics[width=\linewidth]{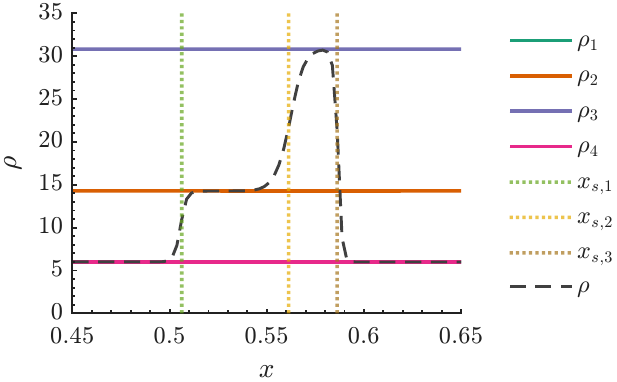}
        \caption{Toro, $t=0.007$.}
        \label{fig:lsDemoToro}
    \end{subfigure}
    \hfill
    \begin{subfigure}[t]{0.32\linewidth}
        \centering
        \includegraphics[width=\linewidth]{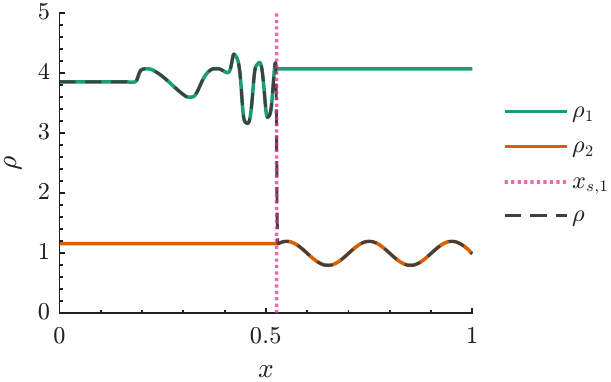}
        \caption{Shu--Osher, $t=0.12$.}
        \label{fig:lsDemoShuOsher}
    \end{subfigure}
    \hfill
    \begin{subfigure}[t]{0.32\linewidth}
        \centering
        \includegraphics[width=\linewidth]{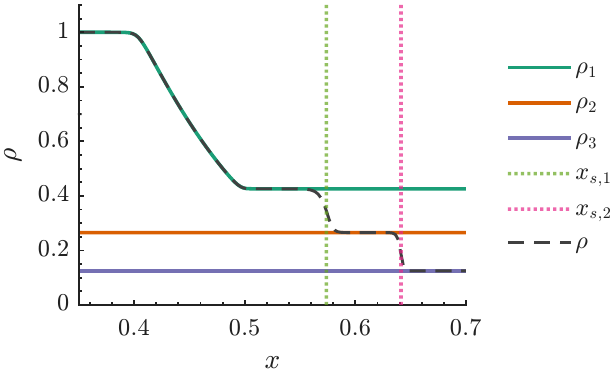}
        \caption{Sod, $t=0.08$.}
        \label{fig:lsDemoSod}
    \end{subfigure}
    \caption{Level set representations of the density fields for the
    one-dimensional Euler test problems: (a) Toro's shock tube problem,
    (b) the Shu-Osher shock-entropy problem, and (c) Sod's shock tube
    problem. The line for the $\rho_1$ component in Toro's shock tube
    problem is hidden by $\rho_4$, as they have the same value.}
    \label{fig:lsDemoEuler}
\end{figure}

Figure~\ref{fig:lsDemoEuler} plots the level set representation of the density
field for each of the demonstration problems described in
Subsection~\ref{ssec:ls1d:description}. We choose to plot the density field
because it varies across the contact surface, while the velocity and pressure
fields are constant. For ease of plotting, we plot the discontinuity location,
rather than its level set function.

\begin{remark}\label{rmk:sharpen}
As discussed in Subsubsection~\ref{sssec:introTanh} and \ref{app:ls1dSharp}, 
compressible flow solvers impose an intrinsic shock width, and preserving this
width helps prevent spurious oscillations. Although we cannot sharpen shocks,
contact surfaces can optionally be sharpened after fitting: we assign them the
smallest fitted shock width in the corresponding flow field. This produces
sharper contacts while retaining a width resolved by the numerical solver. In
our experiments, this choice did not introduce spurious oscillations. This
behavior is consistent with the findings of Fukushima and
Kitamura~\cite{fukushima2024sharp}.
\end{remark}

\subsection{Level Set EnKF Formulation for Compressible Flows}\label{ssec:ls1dEnKF}

As discussed in Remark~\ref{rmk:sharpen}, we wish for our level set
representations to preserve the intrinsic shock width imposed by the
compressible flow solver. Therefore, we wish treat shock and contact widths as a
property of the discretization rather than as a state variable. Thus, we fit
level set representations to each ensemble member such that they share the same
shock and contact widths. To achieve this, we first fit level set
representations to each ensemble member separately, and then use the median
shock width (across ensemble members) as the shared event width. We then refit
the level set representations using this shared prescribed shock width. Details
are presented in \ref{sapp:1d-comp}.

In addition, if a rarefaction wave is present, we align the corresponding state
extensions using the procedure outlined in \ref{ssec:ls1d-affine}, and augment
the latent variables~\eqref{eq:latent} with the affine-transformation
parameters from~\eqref{eq:affine1d}. Finally, we apply the EnKF analysis
step~\eqref{eq:TransformedEnKF_a} to the level set representations, and then
transform back to state space.

\begin{remark}
The initial level set fits for different ensemble members can be computed in
parallel. After computing the ensemble-median width for each primitive field,
the refits can likewise be performed in parallel. This structure improves the
scalability of the level set fitting step with increasing ensemble size.
\end{remark}

\begin{remark}
To reduce computational cost, we could instead infer a shock width from a single
ensemble member. However, this could result in a shock width that is not
representative of the entire ensemble. Thus, we use the median of the ensemble
of widths.
\end{remark}

\begin{remark}\label{rmk:registration1d}
The alignment procedure for the rarefaction wave borrows ideas from image
registration~\cite{brown1992registration}, which seeks to align different data
sets onto a shared coordinate system. Image registration also underlies the
morphing EnKF~\cite{beezley2008registration}.
\end{remark}

\subsection{Results}\label{ssec:1dResults}

We now demonstrate the level set EnKF for 1D compressible flows. We show that
our method preserves sharp features, while reducing RMSE. In addition, the
ensemble spread is similar in magnitude to the RMSE, demonstrating that our
ensemble accurately represents the uncertainty in our state.
Subsubsections~\ref{sec:1dDemoToro}, \ref{sec:1dDemoShuOsher},
and~\ref{sec:1dDemoSod} demonstrate the methodology for Toro' shock tube
problem, the Shu-Osher shock-entropy problem, and Sod's shock tube problem,
respectively.

\subsubsection{Toro's Shock Tube Problem}\label{sec:1dDemoToro}

\begin{figure}
    \centering
    \includegraphics[width=0.65\linewidth]{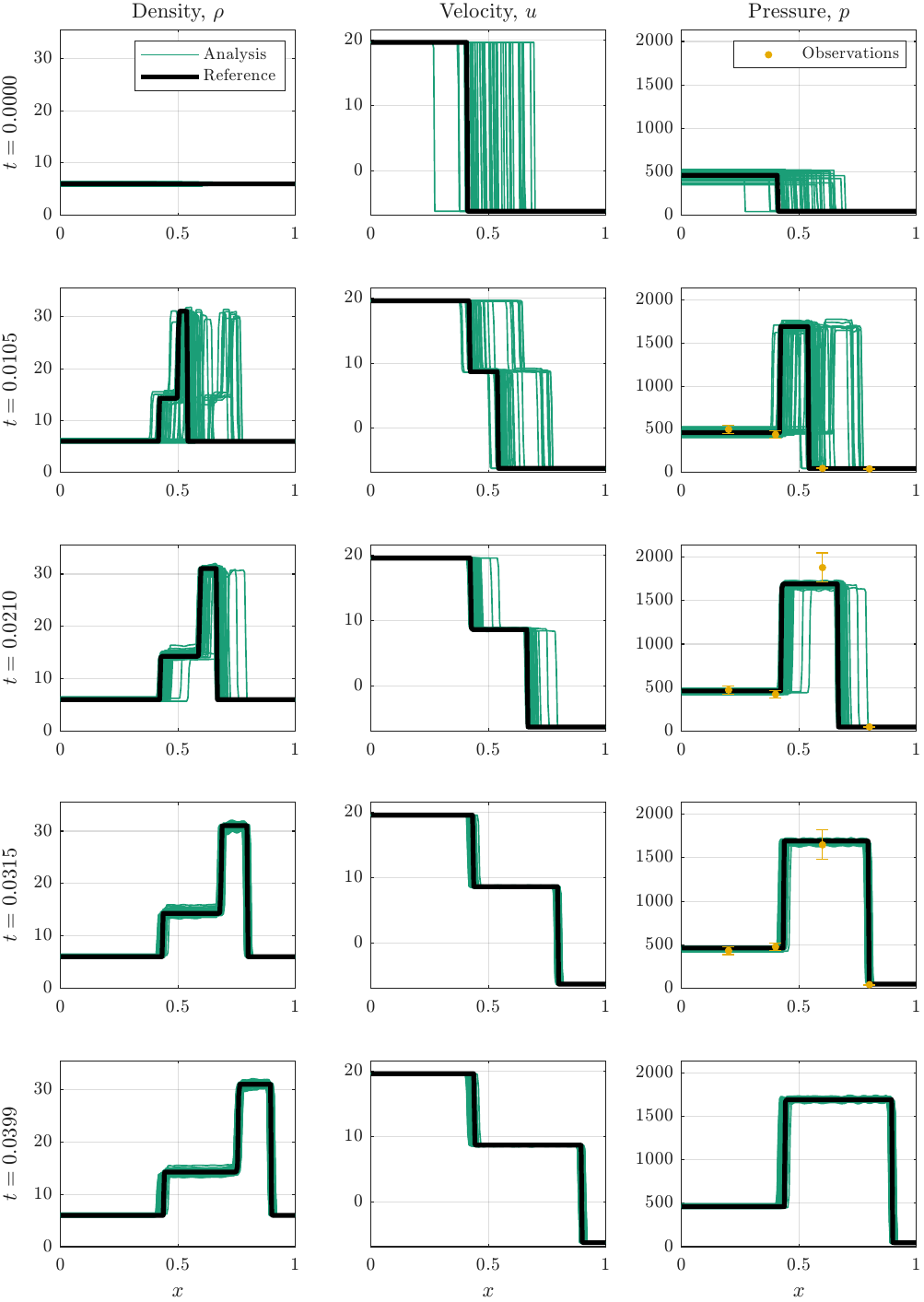}
    \caption{Analysis ensemble for Toro's shock tube problem, plotting the
    ensemble members against the reference truth immediately following the
    analysis step, for a subset of DA cycles. With increasing time, the ensemble
    members show better agreement with the reference truth.}
    \label{fig:toro_da_combined}
\end{figure}

\begin{figure}
    \centering
    \includegraphics[width=0.75\linewidth]{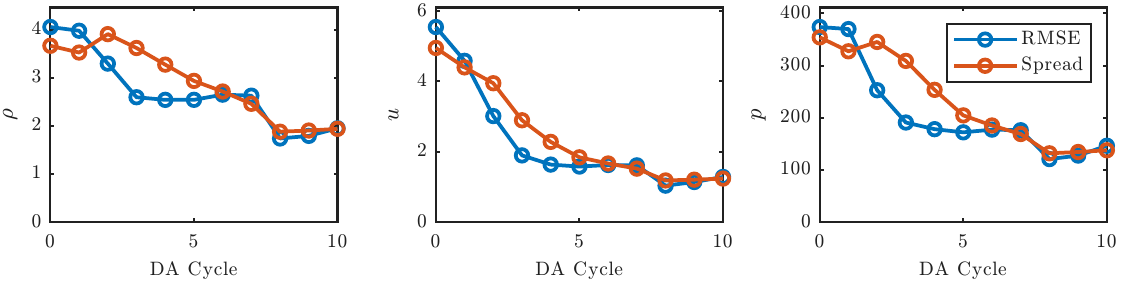}
    \caption{RMSE and spread as a function of DA cycle for Toro's shock tube
    problem. The values at DA cycle 0 correspond to the forecast value before
    the first DA cycle.}
    \label{fig:toro_err}
\end{figure}

\begin{figure}
    \centering
    \includegraphics[width=0.75\linewidth]{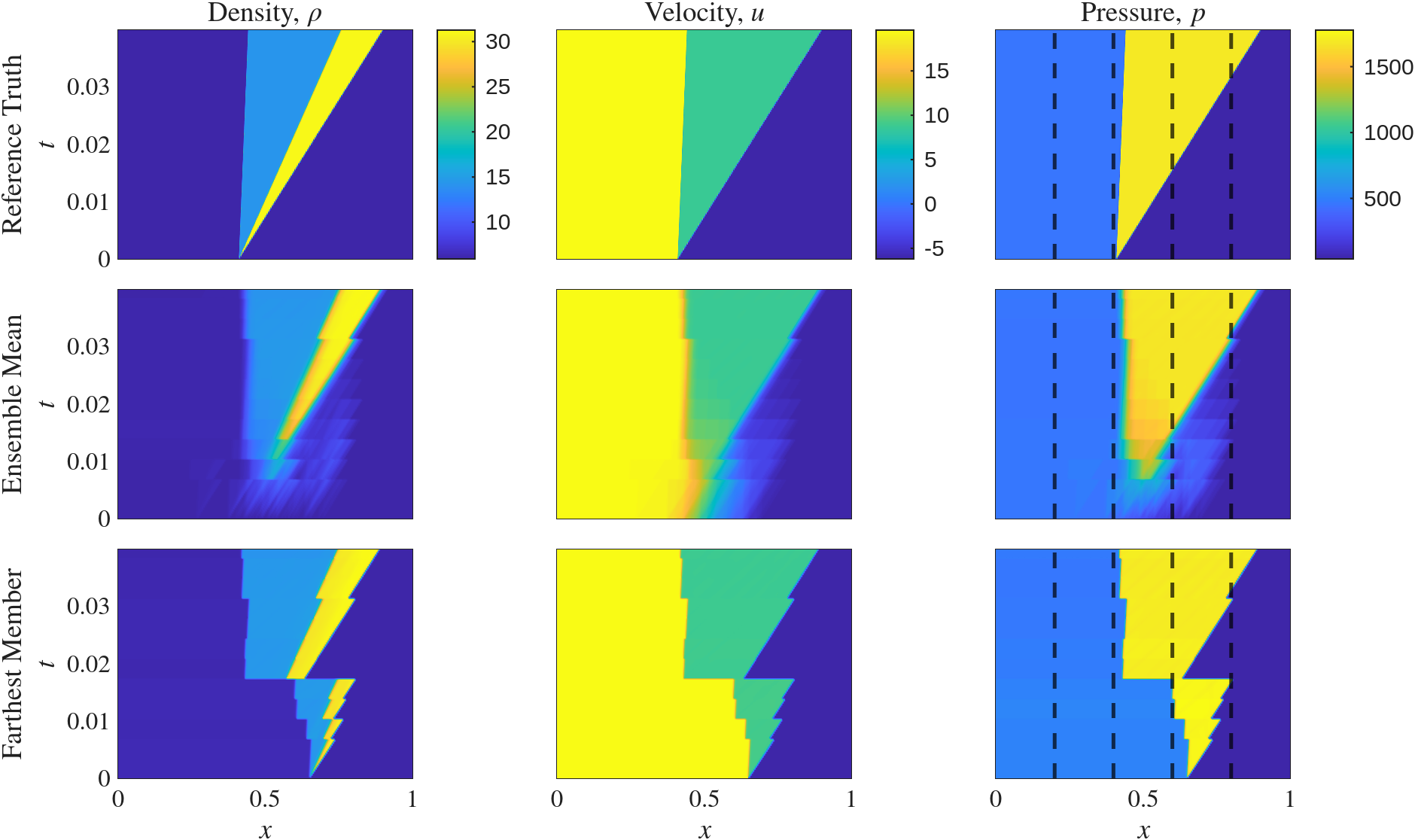}
    \caption{$x-t$ diagrams for Toro's shock tube problem, for the reference
    truth, ensemble mean, and the farthest ensemble member. The dashed vertical
    lines represent the pressure sensors.}
    \label{fig:toro_xt}
\end{figure}

Figure~\ref{fig:toro_da_combined} plots the ensemble members against the
reference truth for the initial condition and several representative analysis
steps. Each analysis step preserves sharp features, while the ensemble converges
to the reference truth as the number of DA cycles increases.
Figure~\ref{fig:toro_err} shows that the RMSE converges and that the ensemble
spread remains comparable in magnitude. The $x$-$t$ diagrams in
Figure~\ref{fig:toro_xt} show that the largest correction occurs when, for a
given member, the right shock reaches the rightmost pressure sensor. The
subsequent update shifts the solution to the left, bringing it to closer
agreement with the reference truth.

\subsubsection{Shu-Osher Shock-Entropy Problem}\label{sec:1dDemoShuOsher}

\begin{figure}
    \centering
    \includegraphics[width=0.65\linewidth]{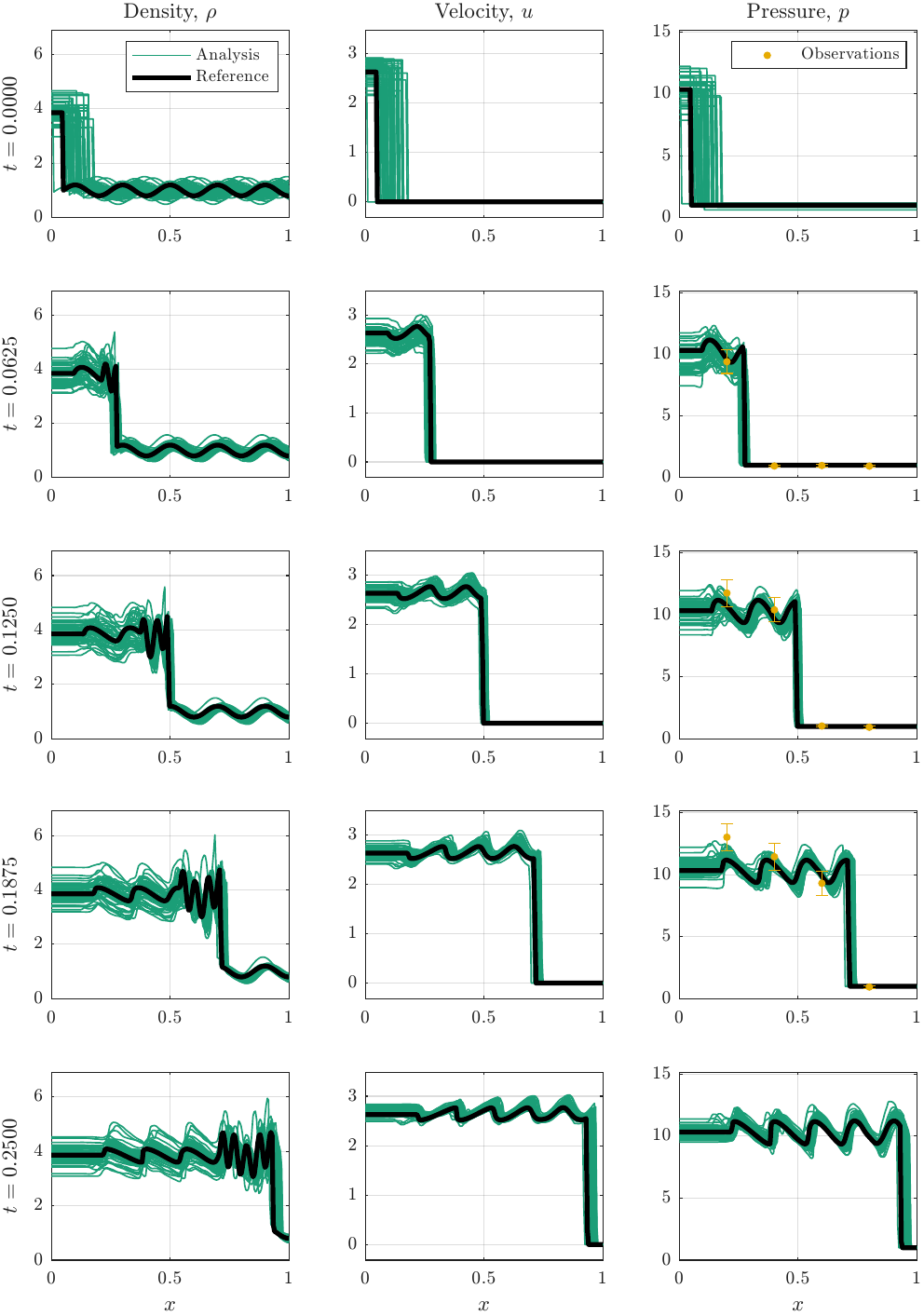}
    \caption{Analysis ensemble for the Shu-Osher shock-entropy problem,
    plotting the ensemble members against the reference truth immediately
    following the analysis step, for a subset of DA cycles. With increasing
    time, the ensemble members show better agreement with the reference truth.}
    \label{fig:shuOsher_da_combined}
\end{figure}

\begin{figure}
    \centering
    \includegraphics[width=0.75\linewidth]{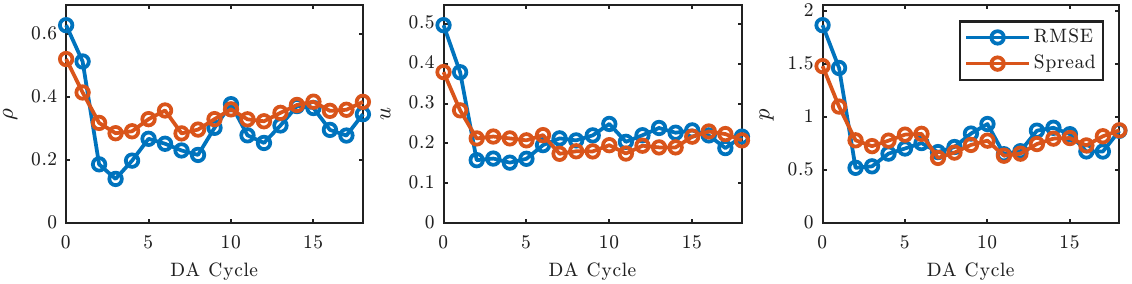}
    \caption{RMSE and spread as a function of DA cycle for the Shu-Osher shock
    entropy problem. The values at DA cycle 0 correspond to the forecast value before
    the first DA cycle.}
    \label{fig:shuOsher_err}
\end{figure}

\begin{figure}
    \centering
    \includegraphics[width=0.75\linewidth]{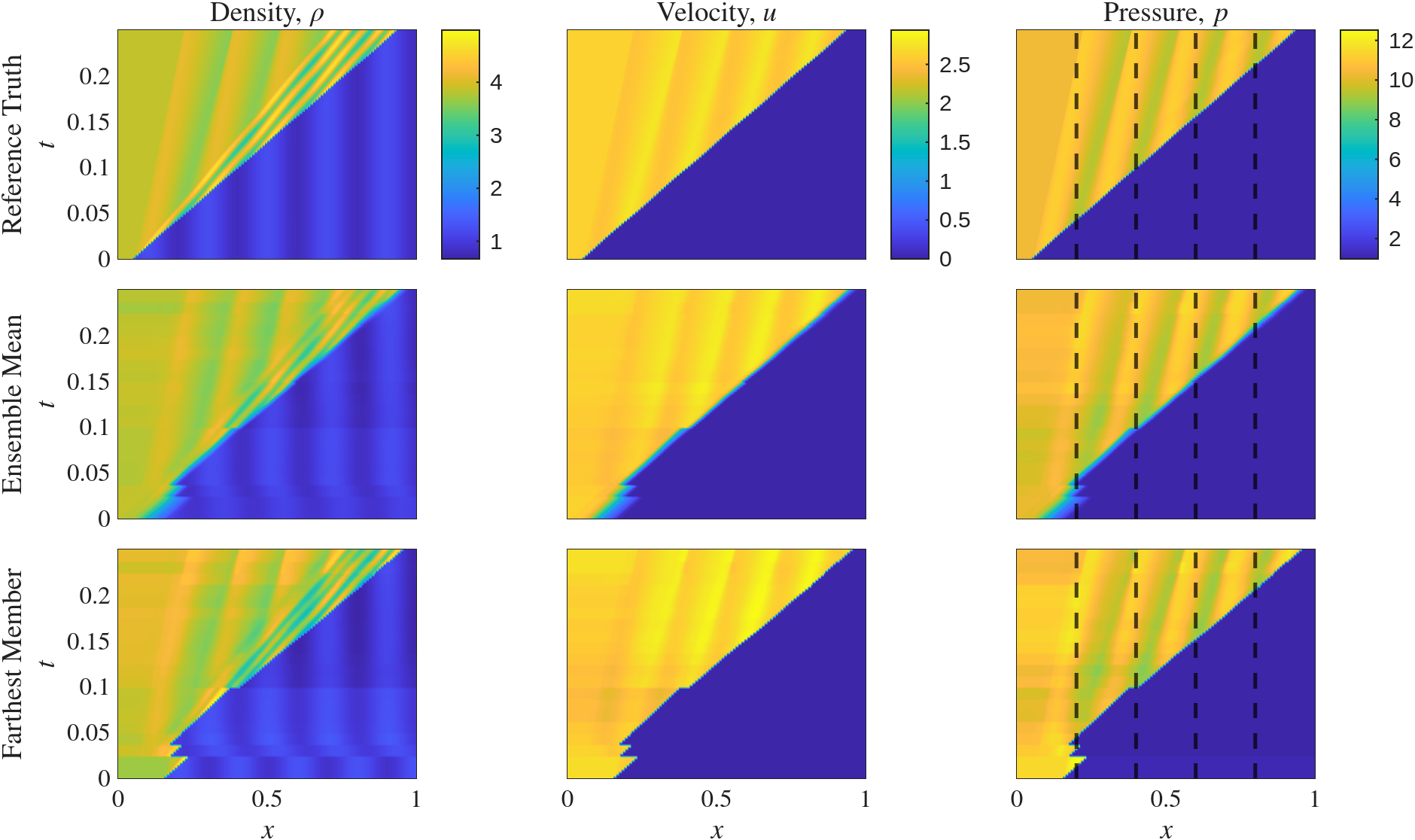}
    \caption{$x-t$ diagrams for the Shu-Osher shock-entropy problem, for the
    reference truth, ensemble mean, and the farthest ensemble member. The
    dashed vertical lines represent the pressure sensors.}
    \label{fig:shuOsher_xt}
\end{figure}

Figure~\ref{fig:shuOsher_da_combined} demonstrates that the level set EnKF
preserves both the sharp shock and the oscillatory features of the entropy wave,
while the ensemble converges to the reference truth as the number of DA cycles
increases. Figures~\ref{fig:shuOsher_err} shows that the ensemble spread remains
a similar magnitude to the RMSE, while the $x$-$t$ diagrams in
Figure~\ref{fig:shuOsher_xt} show improving agreement for both the ensemble mean
and the farthest ensemble member.

\subsubsection{Sod's Shock Tube Problem}\label{sec:1dDemoSod}

\begin{figure}
    \centering
    \includegraphics[width=0.65\linewidth]{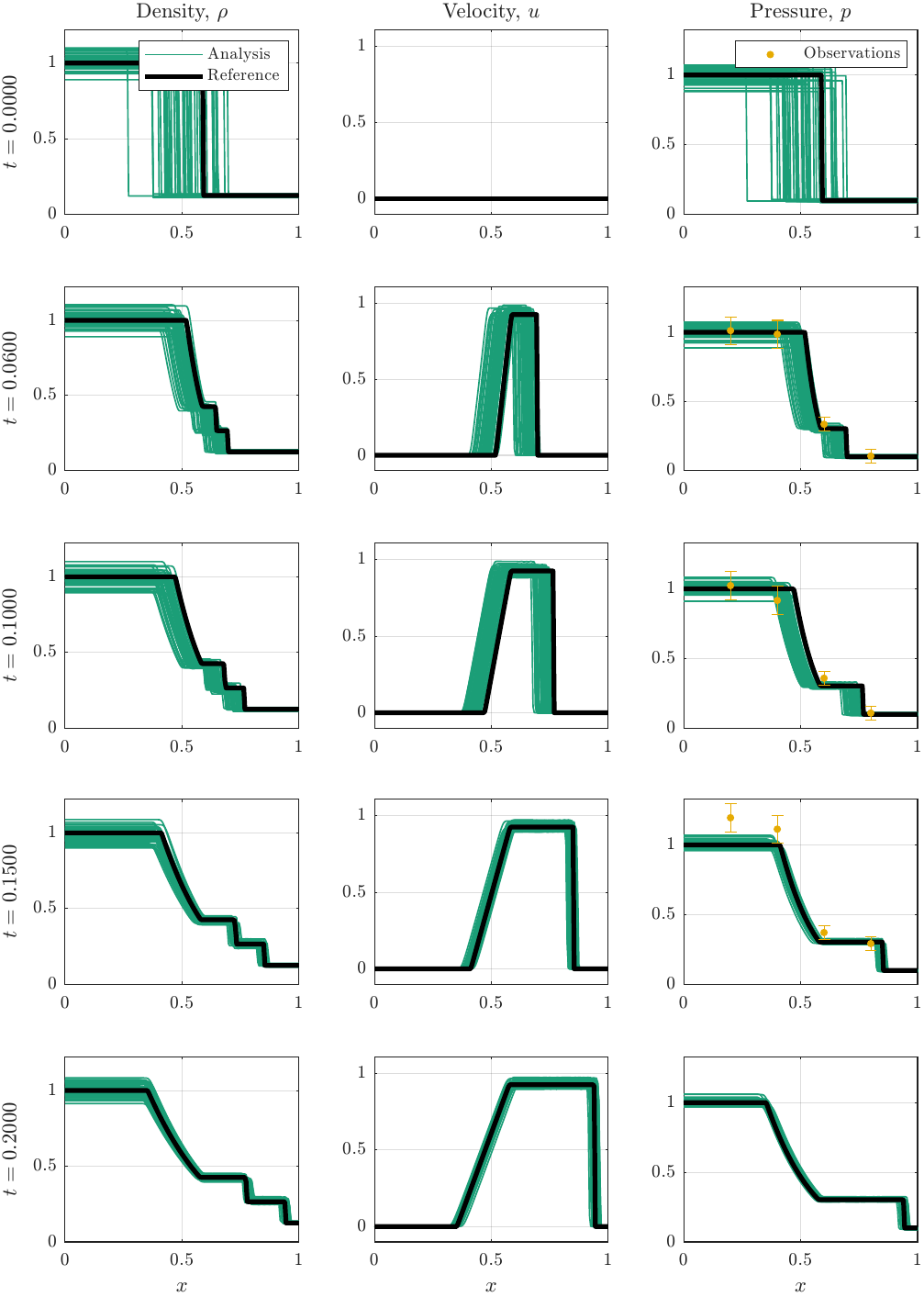}
    \caption{Analysis ensemble for Sod's shock tube problem, plotting the
    ensemble members against the reference truth immediately following the
    analysis step, for a subset of DA cycles. With increasing time, the ensemble
    members show better agreement with the reference truth.}
    \label{fig:sod_da_combined}
\end{figure}

\begin{figure}
    \centering
    \includegraphics[width=0.75\linewidth]{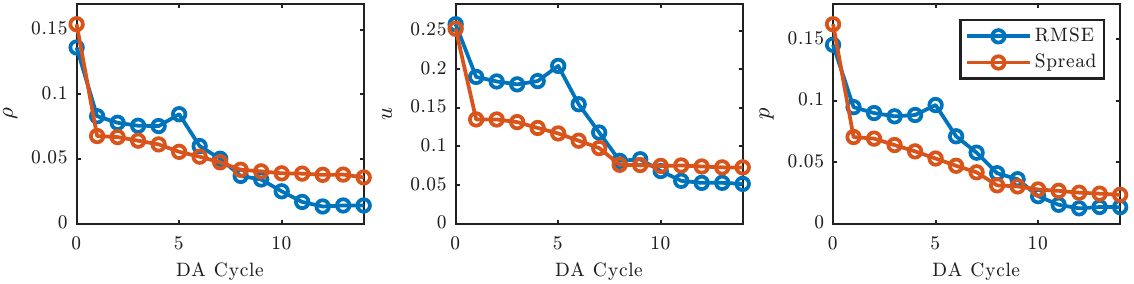}
    \caption{RMSE and spread as a function of DA cycle for Sod's shock tube
    problem. The values at DA cycle 0 correspond to the forecast value before
    the first DA cycle.}
    \label{fig:sod_err}
\end{figure}

\begin{figure}
    \centering
    \includegraphics[width=0.75\linewidth]{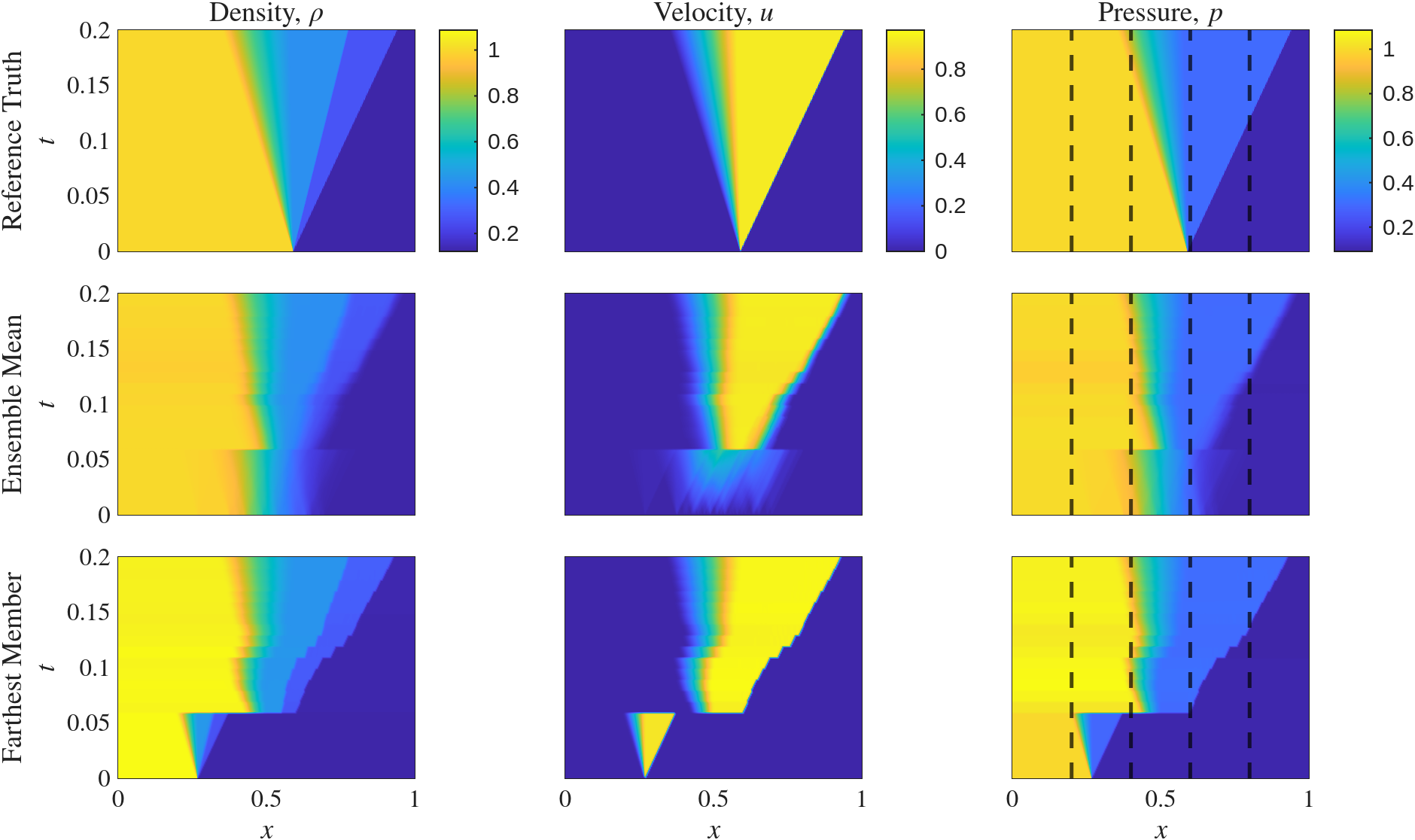}
    \caption{$x-t$ diagrams for Sod's shock tube problem, for the reference
    truth, ensemble mean, and the farthest ensemble member. The dashed vertical
    lines represent the pressure sensors.}
    \label{fig:sod_xt}
\end{figure}

The solution to Sod's shock tube problem comprises a rarefaction wave, a
contact surface, and a shock. We represent the shock and contact surface using
the level set representation, while we align the rarefaction wave using the
affine transformation introduced in \ref{ssec:ls1d-affine}.
Figure~\ref{fig:sod_da_combined} shows that all features are preserved by the
level set EnKF and that the ensemble converges to the reference truth with
increasing DA cycle. Figure~\ref{fig:sod_err} shows that both RMSE and
ensemble spread decrease and remain of similar order. The $x$-$t$ diagrams in
Figure~\ref{fig:sod_xt} show the ensemble mean and farthest ensemble member
approaching the reference solution.

\section{Level Set EnKF: 2D Compressible Euler Equations}\label{sec:ls2dEuler}

Subsection~\ref{ssec:ls2DEulerLS} extends the level set representation to
2D compressible flows, with further details provided in \ref{app:ls2dDetails}.
Subsection~\ref{ssec:ls2dbw} introduces the 2D blast wave problem studied by
Subrahmanya and Sandu~\cite{subrahmanya2025etpfdtw}, and
Subsection~\ref{ssec:enkf2d} demonstrates the level set EnKF for this problem.

\subsection{2D Level Set Representation}\label{ssec:ls2DEulerLS}

In 1D, a discontinuity is located at a single point, whereas in 2D it is
represented by a curve. Let $\Gamma$ denote such a curve. The corresponding 2D
level set function can be defined as the signed distance function
\begin{equation}\label{eq:ls2Dsd}
    \varphi(\bm{x}) =
    \begin{cases}
        \operatorname{dist}(\bm{x},\Gamma),
        & \bm{x} \in \Omega_+,\\
        -\operatorname{dist}(\bm{x},\Gamma),
        & \bm{x} \in \Omega_-,
    \end{cases}
\end{equation}
where $\Omega_-$ denotes the region enclosed by $\Gamma$, and $\Omega_+$
denotes the exterior region. Note that the zero level set of $\varphi(\cdot)$
defines the curve $\Gamma$. Further details on identifying the zero level set
in two dimensions are provided in \ref{sapp:ls2dDetails-Zero}.

State extensions are constructed by applying the 1D methodology in the direction
normal to the discontinuity. Details of this construction for 2D scalar-valued
functions are provided in \ref{sapp:ls2dRep}, and its extension to the 2D
compressible Euler equations is presented in \ref{sapp:ls2dApplyEuler}.

\begin{remark}\label{rmk:fit2D}
As discussed in Remark~\ref{rmk:sharpen}, we preserve the shock width imposed by
the compressible flow solver and treat the fitted width as a property of the
discretization rather than as a state variable. In 1D, we first fit each
ensemble member independently. For each primitive field, we take the minimum
fitted event width within each member, compute the median of those widths across
members, and refit that field in every member with the median width held fixed.
In 2D, we first fit the normal-direction profiles for each primitive field in
the first ensemble member. We take the median of their fitted widths and refit
those profiles, using the median as a prescribed width. We then use the same
prescribed width when fitting the remaining ensemble members.
\end{remark}

\begin{figure}
    \centering
    \includegraphics[width=0.45\linewidth]{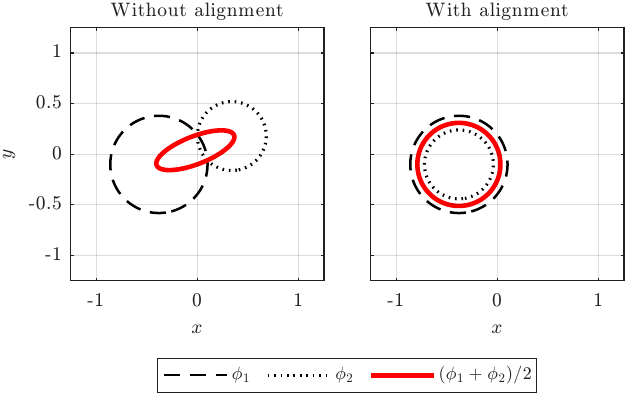}
    \caption{Mean of two circular level set functions, $\phi_1$ and $\phi_2$
    with misaligned centers. The red contours show the zero level sets of
    $(\phi_1+\phi_2)/2$. Without alignment, zero level set of the mean is an
    ellipse. With alignment, it is a circle. Thus, alignment preserves the shape
    of the original discontinuity curve.}
    \label{fig:b2d_ls_alignment}
\end{figure}

In 1D, the level set function measures the signed distance to a set of ordered
points, whereas in 2D, it encodes the signed distance to curves. If these curves
are not aligned, then the signed distance measured by $\varphi$ will not have a
consistent meaning across ensemble members. To demonstrate this, we take the
mean of the level set functions $\varphi_1$ and $\varphi_2$, corresponding to
circles whose centers are not aligned. Figure~\ref{fig:b2d_ls_alignment} shows
that the resulting zero level set is not generally a circle if the circles are
not aligned, while aligning them resolves this issue. Therefore, we use a
registration procedure, described in \ref{ssapp:ls2DregistrationLS}, to align
the level set functions. Additionally, as in the 1D case, we align rarefaction
waves, as described in \ref{ssapp:ls2DregistrationLF}. We augment the 2D level
set representations with the alignment parameters when we apply the transformed
EnKF~\eqref{eq:TransformedEnKF_a}.

\begin{remark}
Remark~\ref{rmk:registration1d}, we explained that we borrow ideas from image
registration~\cite{brown1992registration} to align rarefaction waves in 1D. In
2D, we also use these ideas to align the level set functions. We reiterate that
image registration also underlies the morphing
EnKF~\cite{beezley2008registration}.
\end{remark}

\subsection{2D Blast Wave Problem Description}\label{ssec:ls2dbw}

For the square domain $\Omega\equiv[0,2]\times[0,2]$, we introduce the 2D
blast wave problem with initial condition
\begin{subequations}\label{eq:riemannProblem2d}
\begin{equation}
    \bm{w}(x)=
    \begin{cases}
        \bm{w}_1 & (x,y)\in\Omega_1,\\
        \bm{w}_2 & (x,y)\in\Omega_2.
    \end{cases}
\end{equation}
for
\begin{equation}
    \Omega_1\equiv\{\bm{x}\in\Omega | \|\bm{x}-\bm{x}_0\|_2 < r\},\quad
    \Omega_2\equiv\Omega\setminus\Omega_1,
\end{equation}
where $\bm{x}\in\mathbb{R}^2$.
\end{subequations}
We prescribe the interior and exterior states $\bm{w}_1$ and $\bm{w}_2$,
in terms of the primitive variables $\bm{w}=(\rho,u,v,p)$. Their values, along
with the radius $r$ and center $\bm{x}_0$ of region $\Omega_1$ are specified in
Table~\ref{tab:2d-euler-reference-parameters}.

\begin{figure}
    \centering
    \begin{subfigure}[t]{0.47\linewidth}
        \centering
        \includegraphics[width=\linewidth]{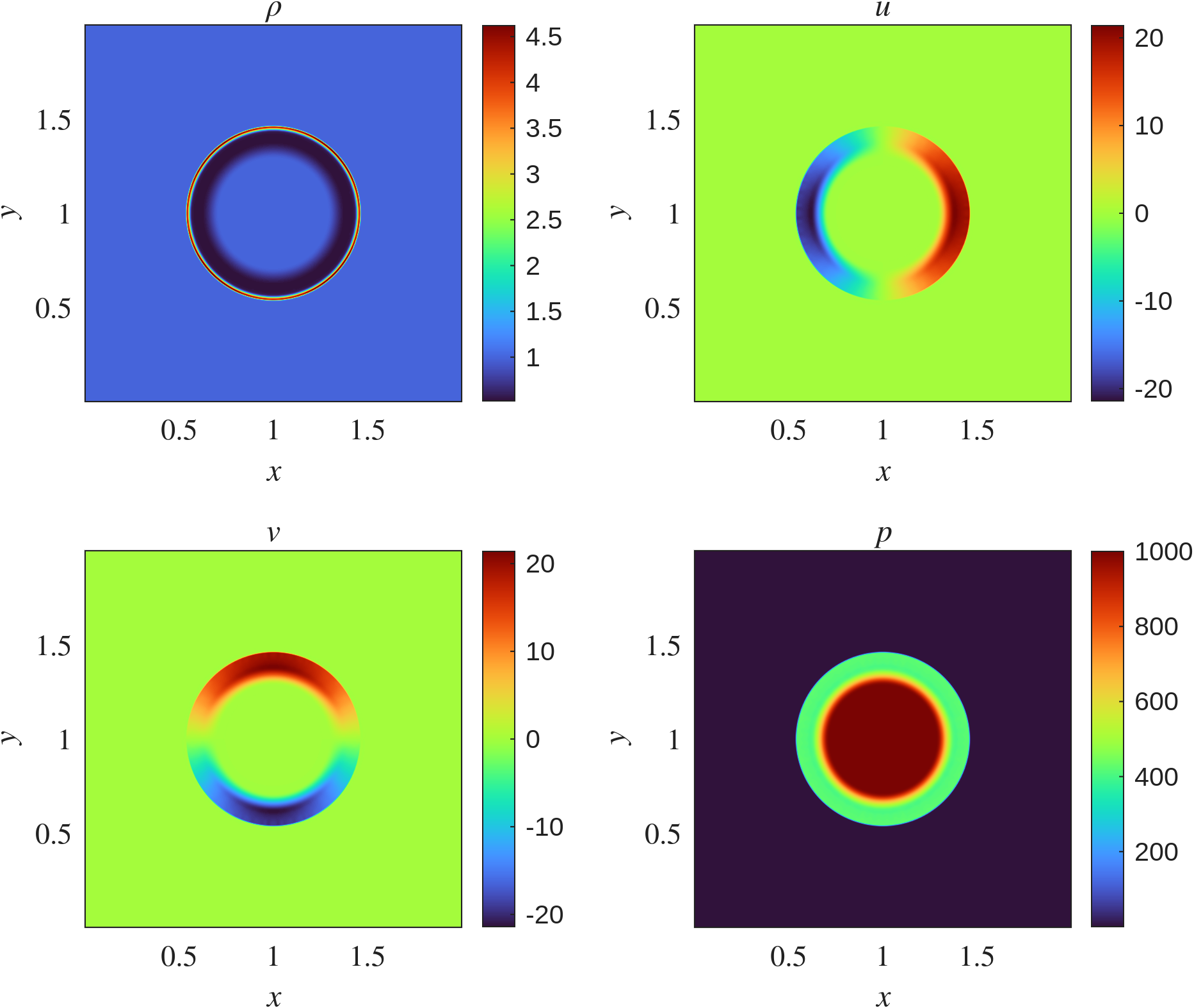}
        \caption{Solution fields.}
        \label{fig:bw2dStates}
    \end{subfigure}
    \hfill
    \begin{subfigure}[t]{0.47\linewidth}
        \centering
        \includegraphics[width=\linewidth]{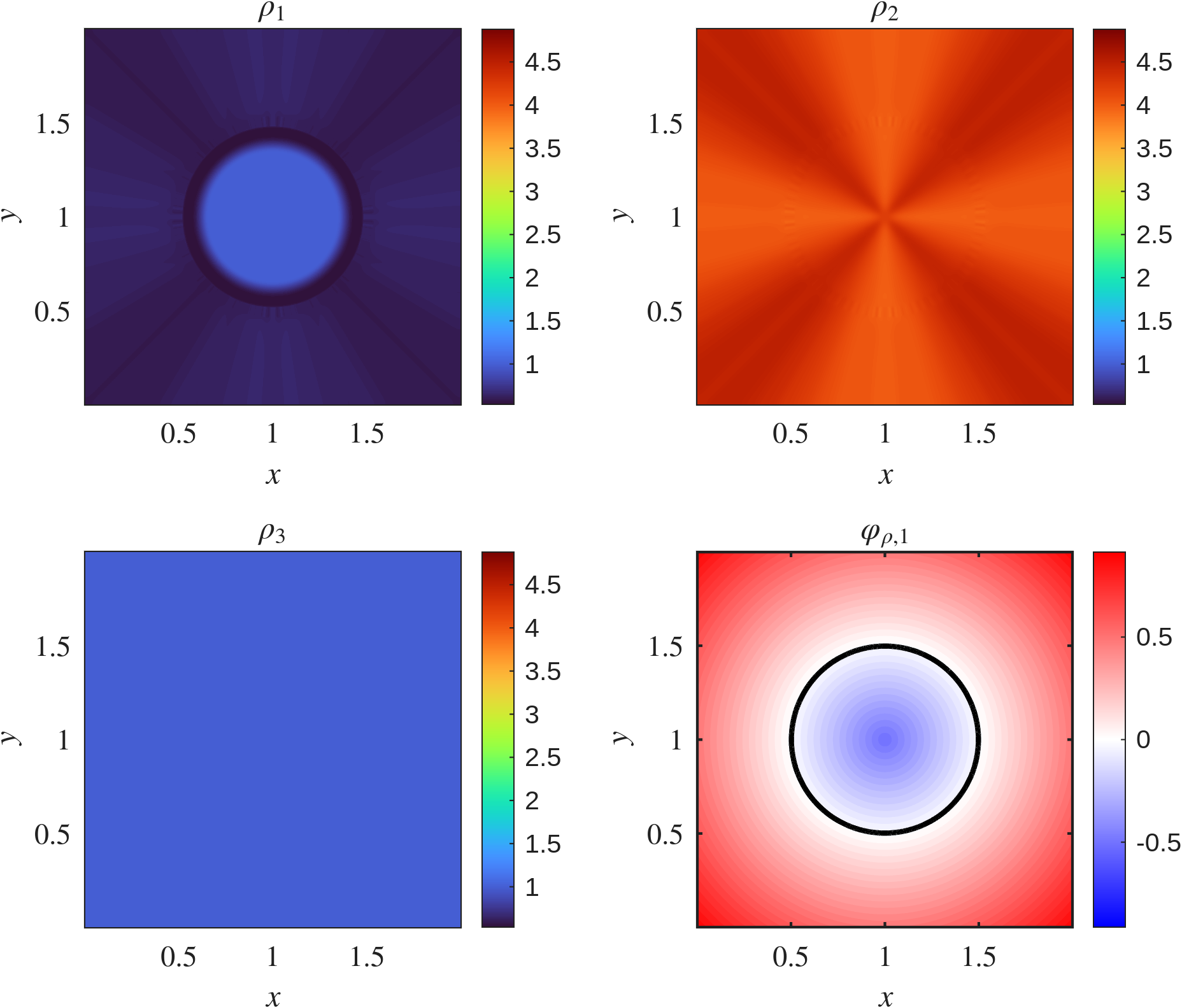}
        \caption{Density level set representation.}
        \label{fig:bw2dLSRho}
    \end{subfigure}
    \caption{Solution and level set representation for the blast wave problem
    at $T=0.025$ time units. In (b), the top-left panel shows the region
    associated with the rarefaction wave, the top-right panel shows the region
    between the contact and the shock, and the bottom-left panel shows the
    region outside the shock. The bottom-right panel shows the level set
    function associated with the contact surface; the black line is its zero
    level set and therefore identifies the contact-surface location.}
    \label{fig:bw2dSolutionLS}
\end{figure}

We time march the reference truth solution to $T=0.025$, depicted in
Figure~\ref{fig:bw2dStates}, and compute its level set representation.
Figure~\ref{fig:bw2dLSRho} shows the resulting level set representation of the
density field.

\begin{table}
    \centering
    \small
    \renewcommand{\arraystretch}{1.25}
    \setlength{\tabcolsep}{5pt}
    \begin{tabular}{@{}l|ccc|cccccc@{}}
        \hline
        Statistic &
         $x_0$ & $y_0$ & $r$ &
         $\rho_1$ & $\rho_2$ & $u_1$ & $u_2$ & $p_1$ & $p_2$ \\
        \hline
        Truth
        & $1.05$ & $1.05$ & $0.45$ & $1.0$ & $1.0$ & $0$ & $0$ & $1000.0$ & $0.01$ \\
        $\mu$
        & $1.0$ & $1.0$ & $r^\dagger$
        & $\rho_1^\dagger$ & $\rho_2^\dagger$
        & $u_1^\dagger$ & $u_2^\dagger$
        & $p_1^\dagger$ & $p_2^\dagger$ \\
        $\sigma$
        & $0.08$ & $0.08$ & $0.05$ & $0.05$ & $0.05$ & $0$ & $0$ & $100.0$ & $0.001$ \\
        \hline
    \end{tabular}
    \caption{Truth parameters, together with the means and standard deviations
    used to construct the initial ensemble for the 2D blast wave experiment.}
    \label{tab:2d-euler-reference-parameters}
\end{table}

\subsection{Level Set EnKF for the 2D Blast Wave Problem}\label{ssec:enkf2d}

As in our 1D examples, uncertainty in the initial condition leads to uncertainty
in our state estimate, which we reduce using the level set EnKF. We again assume
that the parameters specifying the initial condition are known within error
tolerances prescribed in Table~\ref{tab:2d-euler-reference-parameters}, and we
draw each parameter according to~\eqref{eq:randomVar}. Note that the ensemble
means match the truth values for the inner and outer states, while the center of
region $\Omega_1$ is offset from the truth.

\begin{figure}
    \centering
    \includegraphics[width=0.3\linewidth]{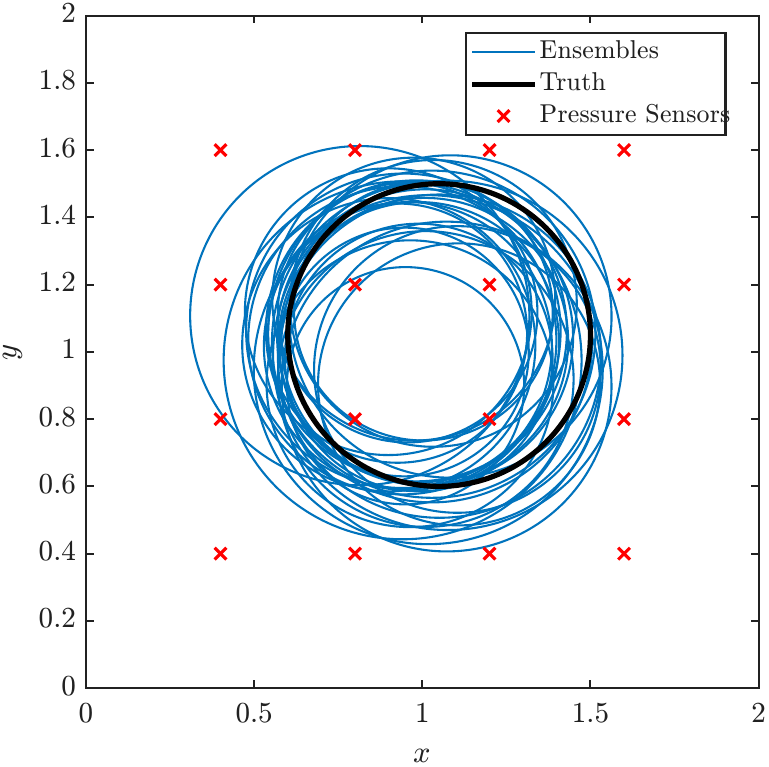}
    \caption{Ensemble of sampled initial discontinuity locations obtained for
    the blast wave problem, overlayed with the initial discontinuity location
    for the reference truth and the sensor locations.}
    \label{fig:bw2d_IC}
\end{figure}

We generate pressure observations from the reference solution as
\begin{equation}
    d_i = p_{\mathrm{ref}}(x_{\mathrm{obs},i},y_{\mathrm{obs},i},t;
    \bm{q}_1,\bm{q}_2,x_0,y_0,r)+\eta_i,\quad i=0,\dots,N_d-1,
\end{equation}
for $\bm{d}\in\mathbb{R}^{N_d}$, where $\bm{\eta}\sim\mathcal{N}(0,R)$ for the
observation-error covariance $R$, whose diagonal entries are given by
$R_{ii}=\max\{0.1p_{\mathrm{ref}}(x_{\mathrm{obs},i},y_{\mathrm{obs},i},t;
\bm{q}_1,\bm{q}_2,x_0,y_0,r),0.05\}$. This choice imposes a relative observation
error of 10\% and and an absolute error of 0.05. We observe pressure on the
uniform $4\times4$ grid so that the full observation set is given by the
tensor-product grid
\[
(\bm{x}_{\mathrm{obs}},\bm{y}_{\mathrm{obs}})
\in
\{0.4,0.8,1.2,1.6\}
\times
\{0.4,0.8,1.2,1.6\}.
\]
Thus, each assimilation cycle uses 16 pressure observations.
Figure~\ref{fig:bw2d_IC} depicts $\Omega_1$ for each ensemble against the
reference truth, as well as the pressure sensors.

\begin{figure}
    \centering
    \includegraphics[width=0.95\linewidth]{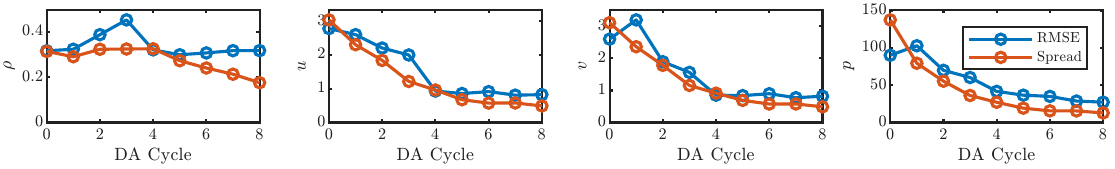}
    \caption{RMSE and spread as a function of DA cycle for the blast wave
    problem. The density is relatively constant throughout the domain, so the
    RMSE and spread are both small throughout the whole calculation. The spread
    and RMSE in the velocity and pressure fields is initially larger, and
    decreases as more data is assimilated. The spread and RMSE have similar
    orders of magnitude throughout the calculation. The values at DA cycle 0
    correspond to the forecast value before the first DA cycle.}
    \label{fig:bw2d_err}
\end{figure}

The simulation is advanced to $T=0.01$, with the first observation arriving at
$t=0.0025$ and subsequent DA cycles every $\Delta t_{\mathrm{DA}}=0.001$.
Figure~\ref{fig:bw2d_err} shows the RMSE and ensemble spread as functions of DA
cycle. As pressure observations are assimilated, both the RMSE and ensemble
spread decrease, indicating that the filter reduces state uncertainty while
maintaining an ensemble spread of the same order as the actual error. The
density RMSE and spread remain small throughout the assimilation window because
the density field is nearly uniform over the domain.

\begin{figure}
    \centering
    \includegraphics[width=0.8\linewidth]{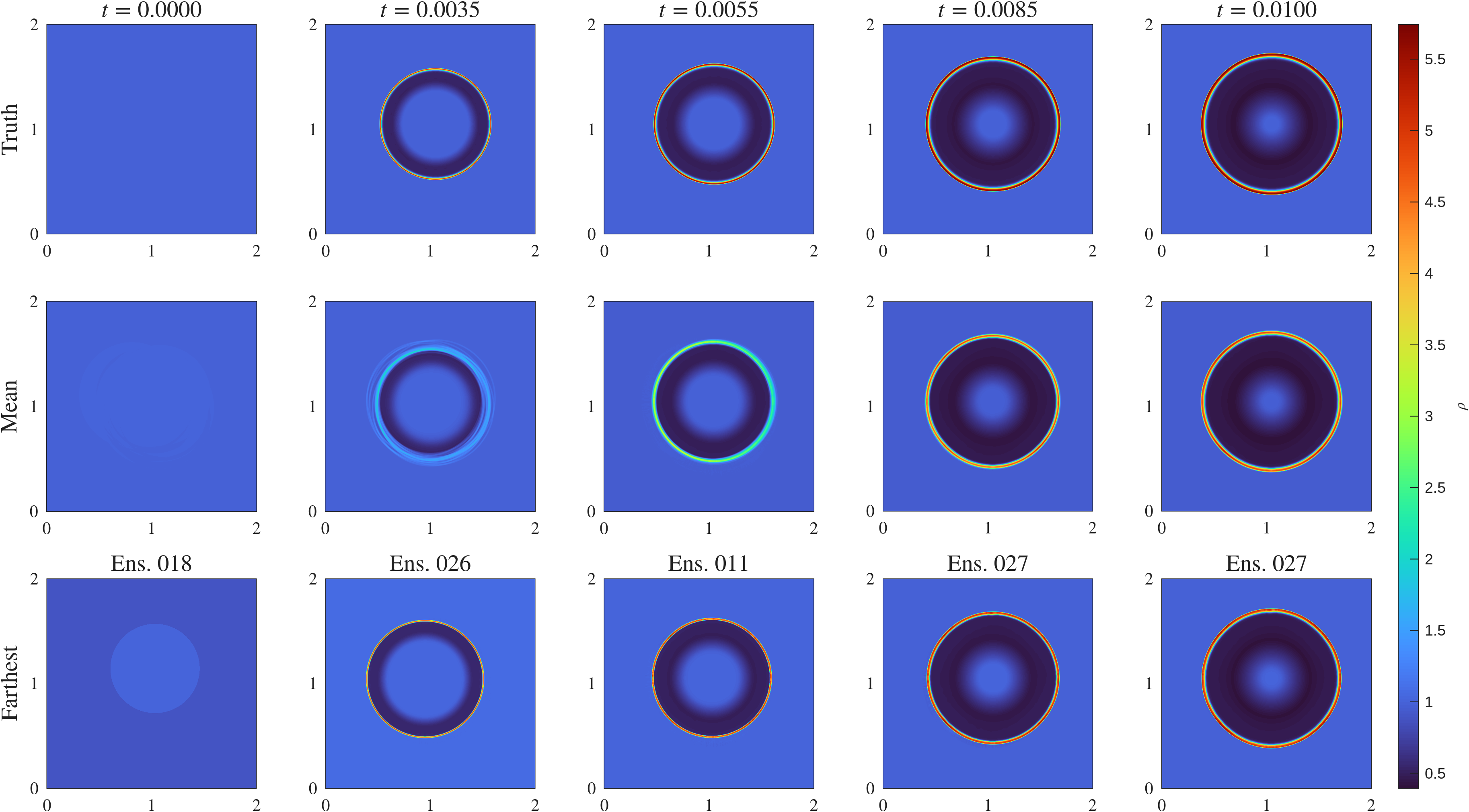}
    \caption{Density field evolution for the 2D blast-wave DA experiment.
    Columns show selected DA cycles from the initial condition through the final
    cycle; rows show the truth, ensemble mean, and the farther ensemble member.}
    \label{fig:bw2d_DA_rho}
\end{figure}

\begin{figure}
    \centering
    \includegraphics[width=0.8\linewidth]{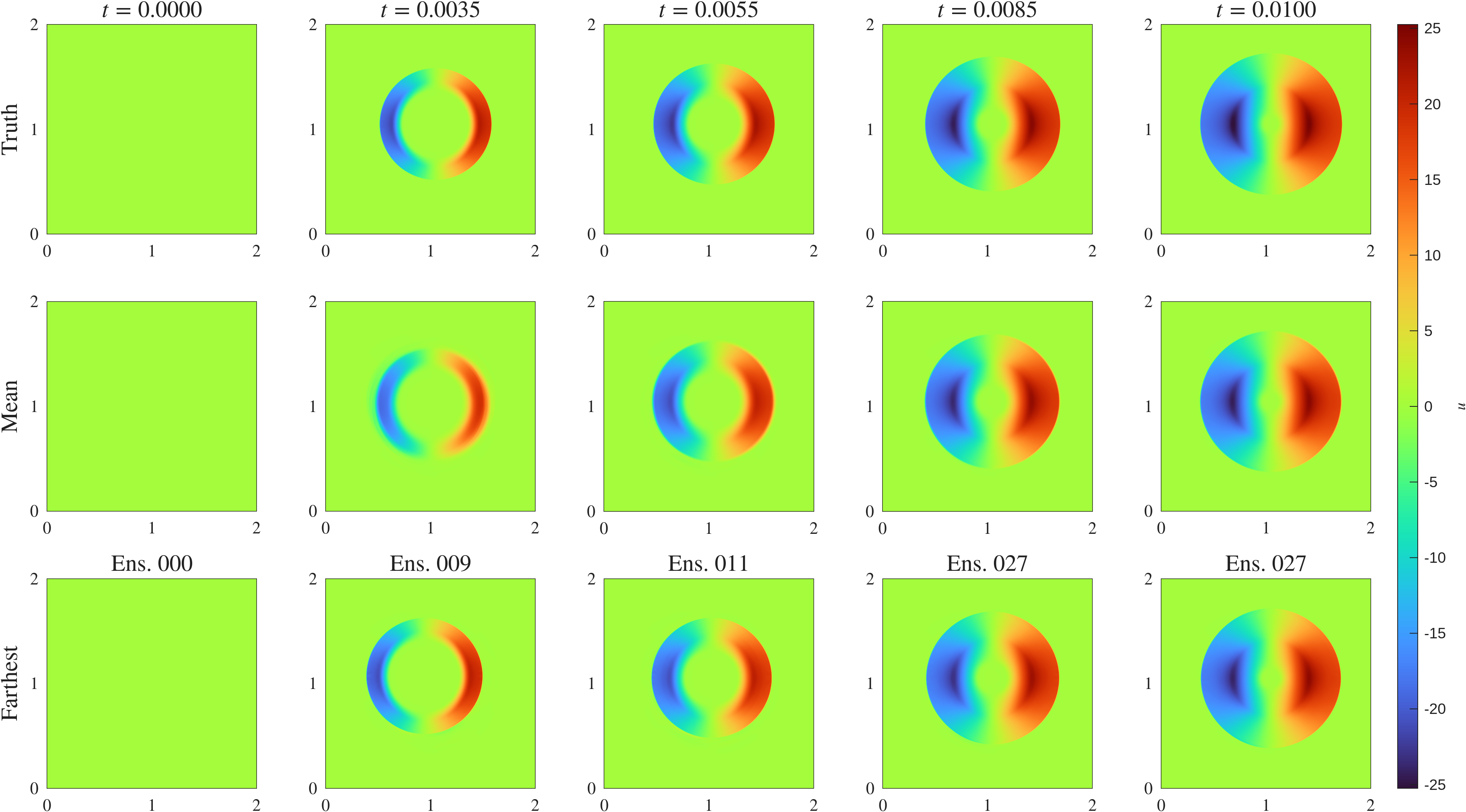}
    \caption{$u$-velocity field evolution for the 2D blast-wave DA experiment.
    Columns show selected DA cycles from the initial condition through the final
    cycle; rows show the truth, ensemble mean, and the farther ensemble member.}
    \label{fig:bw2d_DA_u}
\end{figure}

\begin{figure}
    \centering
    \includegraphics[width=0.8\linewidth]{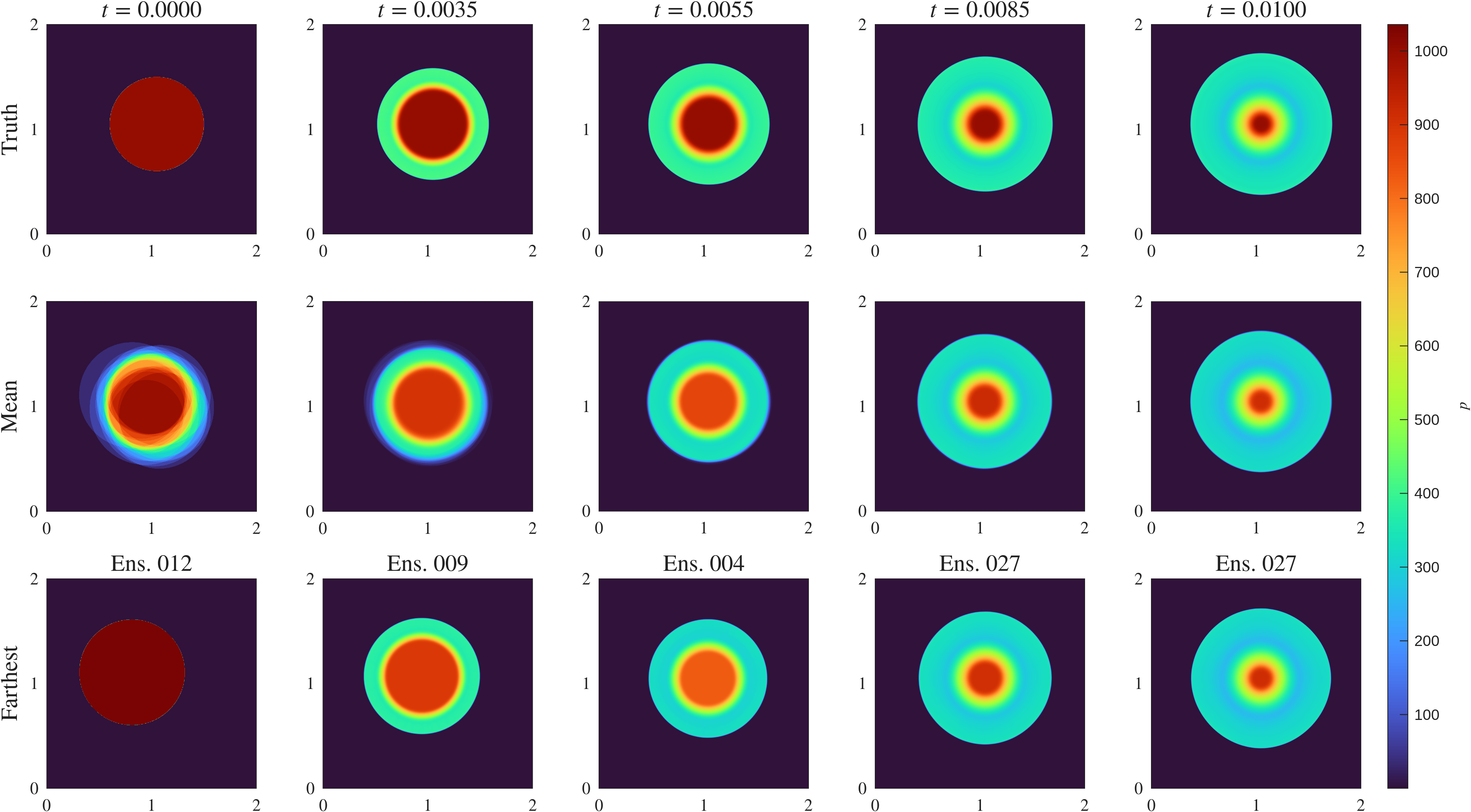}
    \caption{Pressure field evolution for the 2D blast-wave DA experiment.
    Columns show selected DA cycles from the initial condition through the final
    cycle; rows show the truth, ensemble mean, and the farther ensemble member.}
    \label{fig:bw2d_DA_p}
\end{figure}

Figures~\ref{fig:bw2d_DA_rho}, \ref{fig:bw2d_DA_u}, and~\ref{fig:bw2d_DA_p} plot
contours of the density, $u$-velocity, and pressure, respectively. For each
variable, the first row plots the truth, the second row plots the ensemble mean,
and the third row plots the farthest ensemble member. We define the farthest ensemble
member as described in Subsubsection~\ref{sssec:setup1dMetrics}. For all variables, we
observe that the farthest ensemble member moves closer to the reference truth as
more data is assimilated, so that the ensemble mean converges to the truth.

\section{Conclusion}\label{sec:conclusion}

This paper introduces the \textit{level set EnKF}, which enables sequential DA
in compressible flows with shocks and other sharp discontinuities. In this
approach, the filter is applied in smooth latent variables, rather than the
original state space. The latent variables comprise a level set function that
encodes discontinuity location and smooth, globally defined, state extensions,
that represent the solution on either side of the discontinuity. By applying
the EnKF analysis step in this latent space, rather than directly in state
space, the method avoids taking linear combinations of ensemble members with
misaligned discontinuities, which is the underlying cause of spurious
oscillations introduced by the standard EnKF for these flows.

The level set EnKF is limited by its ability to accurately detect
discontinuities, making it less flexible than the morphing
EnKF~\cite{beezley2008registration} and the neural-network-based
approaches~\cite{zhou2026neuralenkf,chandravamsi2026decoder}. However, its
latent space representation is physically interpretable, which is useful for
both understanding the action of the filter and incorporating problem-specific
inductive bias. In addition, the level set representation explicitly partitions
the physical domain into regions associated with distinct flow features. This
separation prevents the analysis step from directly coupling regions whose
uncertainties are largely unrelated, resonating with other localization
techniques that are important in the use of the EnKF \cite{asch2016data}.
Although the present methodology is limited to relatively simple geometries, it
may offer advantages over competing methods when shocks and other sharp
interfaces provide a geometrically simple partition of the flow.

More broadly, the level set representation is closely related to methods for
shock detection, which remains a challenging post-processing problem in
compressible flows. In particular, the methodology introduced in this paper
could be used to quantify both shock strength and shock location, as it
explicitly identifies a shock location and function values on either side of the
shock. Moreover, other shock detection methods could be used to enhance the
methodology presented in this paper.

Several limitations remain. The present implementation assumes that the number
of discontinuities is known in advance and that each discontinuity can be
represented using a hyperbolic tangent function. Extending the method to
problems involving more complicated geometries, shock interactions, and
intersecting discontinuities will require more flexible procedures for
automatic feature identification and latent-space construction. Future work
should also investigate more flexible, potentially learned representations of
discontinuities as alternatives to the hyperbolic tangent function. Finally, as
discussed in Remark~\ref{rmk:jacobian}, more efficient methods for computing or
approximating the Jacobians used in the nonlinear least-squares problem should
be explored to improve scalability to large-scale problems. These directions are
natural next steps toward applying feature-preserving DA to more complex
compressible-flow problems.

\section*{Code availability}

All code used to generate the results and figures presented in this paper is
publicly available at
\href{https://github.com/mkesleem/LevelSetEnKF}
{https://github.com/mkesleem/LevelSetEnKF}.

\section*{Data availability}

Data will be made available on request.

\section*{Declaration of generative AI use}

Codex was used to improve the efficiency and parallelism of the code for
computing the level set representation, improve the readability of the
manuscript figures, and create tables for the manuscript. The authors reviewed
and verified all AI-assisted changes and take full responsibility for the
accuracy, originality, and integrity of the manuscript.

\section*{Declaration of competing interest}

The authors declare that they have no known competing financial interests or
personal relationships that could have appeared to influence the work reported
in this paper.

\section*{Acknowledgments}

This work was supported by the AFOSR/AFRL Center of Excellence in Assimilation
of Flow Features in Compressible Reacting Flows (Grant No. FA9550-25-1-0011).
Matthias Morzfeld is also supported by the U.S. ONR Grant N000142512298.

\bibliographystyle{elsarticle-num} 
\bibliography{references}
\appendix
\section{Compressible Euler Equations}~\label{app:eulerEquations}

This appendix briefly reviews the compressible Euler equations, we refer the
reader to~\cite{toro2009compressible} for a detailed description. The
compressible Euler equations are a system of nonlinear hyperbolic conservation
laws that describe the dynamics of a compressible, inviscid flow within the
continuum approximation. Such flows can be expressed in terms of the
\textit{primitive variables} $(\rho,u,v,w,p)$, where $\rho$ denotes the density,
$\bm{u}=(u, v, w)$ represents the velocity field with components in the $x$,
$y$, and $z$ directions, respectively, and $p$ is the pressure. Alternatively,
the flow can be expressed in terms of the \textit{conservative variables}
$(\rho,\rho u,\rho v, \rho w,E)$, where $\rho$ is the density,
$\rho\bm{u}=\rho(u,v,w)$ is the momentum, and $E$ is the total energy
\begin{equation}
    E
    =\rho\Big(e+\frac{\bm{u}\cdot\bm{u}}{2}\Big)
    =\rho\Big(e+\frac{u^2+v^2+w^2}{2}\Big)
\end{equation}
while $e$ is the specific internal energy and $(\bm{u}\cdot\bm{u})/2$ is the
specific kinetic energy. In conservative form, the compressible Euler equations
can be written as
\begin{subequations}
\begin{align}
    \frac{\partial\rho}{\partial t}+\nabla\cdot(\rho\bm{u})&=0,\\
    \frac{\partial(\rho\bm{u})}{\partial t}
    +\nabla\cdot(\rho\bm{u}\otimes\bm{u})
    +\nabla p&=0,\\
    \frac{\partial E}{\partial t}+\nabla\cdot[(E+p)\bm{u}]&=0.
\end{align}
\end{subequations}
We relate the thermodynamic variables using the ideal gas law
\begin{equation}
    p=\rho R T
\end{equation}
while for thermally and calorically perfect gases, we obtain
\begin{equation}
    e=c_v T
\end{equation}
where $R=c_p-c_v$ is the specific gas constant, $T$ is the temperature, $c_p$ is
the specific heat capacity at constant pressure, and $c_v$ is the specific heat
capacity at constant volume. Thus, the pressure can be expressed in terms of the
energy and momentum as
\begin{equation}
    p = (\gamma - 1)\Big[E-\frac{(\rho\bm{u})\cdot(\rho\bm{u})}{2\rho}\Big],
\end{equation}
where $\gamma=c_p/c_v$ is the specific heat capacity ratio. We write the
conservative variables as
\begin{subequations}
\label{eq:cons2prim}
\begin{align}
\bm{q} &=
\begin{bmatrix}
q_1 \\ q_2 \\ q_3 \\ q_4 \\ q_5
\end{bmatrix}
=
\begin{bmatrix}
\rho \\ \rho u \\ \rho v \\ \rho w \\ E
\end{bmatrix},
\end{align}
and the primitive variables as
\begin{align}
\bm{w} &=
\begin{bmatrix}
\rho \\ u \\ v \\ w \\ p
\end{bmatrix}
=
\begin{bmatrix}
q_1 \\ q_2 / q_1 \\ q_3 / q_1 \\ q_4 / q_1 \\ (\gamma-1)[q_5 - (q_2^2 + q_3^2
 + q_4^2)/(2 q_1)]
\end{bmatrix},
\end{align}
\end{subequations}
so that~\eqref{eq:cons2prim} describes the conservative-to-primitive
transformation.
\section{Interface Sharpening Using the Level Set Representation}\label{app:ls1dSharp}

Godunov's scheme is a finite volume method for conservation laws. It uses
piecewise-constant reconstruction, resulting in a method that is first-order
accurate in space and time. Consequently, the scheme introduces significant
numerical dissipation. Higher-order methods, such as the MUSCL
scheme~\cite{vanleer1979muscl} and the WENO method~\cite{liu1994weno}, reduce
smearing. MUSCL schemes use piecewise-linear reconstruction and employ
\textit{slope limiters} to prevent spurious numerical
oscillations, at the expense of smearing of sharp interfaces, albeit, to a
lesser degree than Godunov's method~\cite{vanleer1979muscl}. WENO methods use
multiple candidate stencils for solution reconstruction and compute nonlinear
weights that adaptively reduce the influence of stencils containing
discontinuities~\cite{liu1994weno}. WENO methods substantially improve the
resolution of shocks, relative to MUSCL and Godunov schemes, yet still introduce
some numerical dissipation, which causes sharp interfaces to be smeared over a
small number of grid cells.

Our level set representation models shock and contact surfaces using a
hyperbolic tangent function with a finite width $\delta$. In principle, we could
take the limit $\delta\to0$ to sharpen these interfaces in the analysis step.
However, this would interact unfavorably  with the prediction step. In
particular, the experiments reported in this paper use the \texttt{M2C} code,
based on a MUSCL scheme. Introducing sharp discontinuities in the analysis step
would effectively undo the smoothing introduced by the slope limiters, which are
required to suppress spurious numerical oscillations near discontinuities.
Therefore, we fit a finite interface width $\delta$ when constructing our level
set representation such that $\delta$ respects the length scale of the shock
introduced by the numerical solver.

\begin{figure}
\centering
\includegraphics[width=0.65\linewidth]{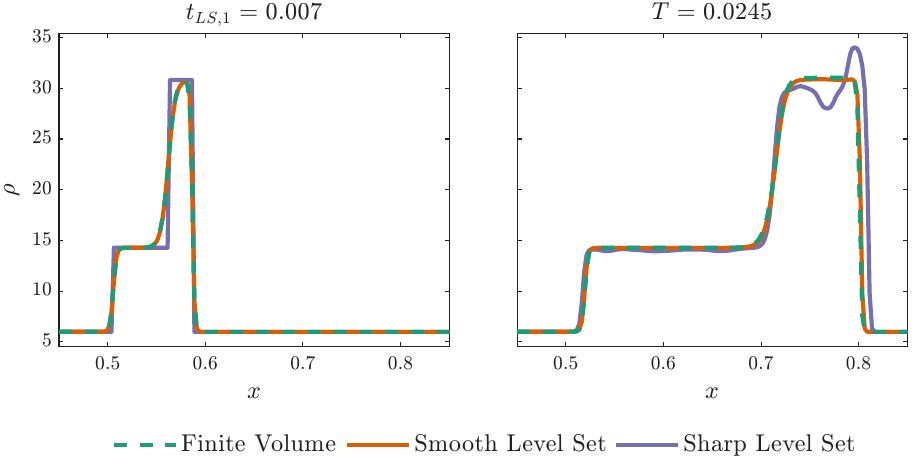}
\caption{Evolution of smooth and sharp level set representations of Toro's
shock tube solution compared with the reference finite volume solution for the
density field.}
\label{fig:toro_LS_Sharp}
\end{figure}

Figure~\ref{fig:toro_LS_Sharp} demonstrates the negative impacts of sharpening
shocks and contact surfaces using the level set representation. We first march
to $t_{LS,1}=0.007$ and fit a level set representation. Starting from this time,
we evolve three solutions: the fitted smooth representation, with $\delta$
determined during fitting; the corresponding sharp representation, obtained by
taking the limit $\delta\to0$; and the reference finite volume solution. After
each additional interval of $\Delta t=0.0035$, the level set representations are
refitted and evolved again. This process is repeated until the final time
$T=0.0245$. We observe that the sharp representation produces larger
oscillations and causes the right-most shock to propagate at an incorrect
speed relative to the reference solution, while the fitted smooth representation
remains much closer to the reference finite volume solution. This behavior is
consistent with the findings of Fukushima and
Kitamura~\cite{fukushima2024sharp}, which we summarize in
Subsubsection~\ref{sssec:introTanh}.

\section{Implementation Details for 1D Level Set Representations}\label{app:ls1dDetails}

This appendix provides implementation details for the level set representation
of 1D scalar fields introduced in Section~\ref{sec:ls1d}.
\ref{sapp:ls1dDetailsLocalProblem} explains how the optimization
problem~\eqref{eq:1dLSObjCont} used to fit the level set reconstruction
map~\eqref{eq:tanh_ls} can be reformulated as a local minimization problem. The
continuous formulation is presented in \ref{sapp:ls1dDetailsLocalProblemCont},
while the corresponding discrete formulation is given in
\ref{sapp:ls1dDetailsLocalProblemDisc}. Algorithms~\ref{alg:1dInit}
and~\ref{alg:1dLevelSetMap}, presented in
\ref{sapp:ls1dDetailsLocalProblemDisc}, together define the procedure used to
construct a level set representation of a one-dimensional function containing a
single discontinuity. Algorithm~\ref{alg:multiDiscontinuityLocalFit} in
\ref{sapp:1dLSDetails-K} describes how to extend the procedure to multiple
discontinuities based on the discussion in Subsection~\ref{ssec:ls1d-K}.
Finally, \ref{sapp:1d-comp} extends the scalar methodology to solutions of the
1D Euler equations.

\subsection{Treatment as a Local Minimization Problem}
\label{sapp:ls1dDetailsLocalProblem}

Although~\eqref{eq:1dLSObjCont} is stated globally, the hyperbolic tangent
transition is localized near the discontinuity. For $x$ sufficiently far to the
left of $x_s$, $\tanh((x-x_s)/\delta)\approx -1$, so
$\mathcal{T}^{-1}(\bm{\phi})(x)\approx f_L(x)$. Similarly, for $x$ sufficiently
far to the right of $x_s$, $\tanh((x-x_s)/\delta)\approx 1$, so
$\mathcal{T}^{-1}(\bm{\phi})(x)\approx f_R(x)$. Thus, the transition from $f_L$
to $f_R$ is confined to a neighborhood of $x_s$, and we can restrict the
minimization problem to this neighborhood. We identify this neighborhood using
the local minima of $|f'(x)|$ to the left and right of the discontinuity.

\subsection{Continuous Formulation of the Local Minimization Problem}
\label{sapp:ls1dDetailsLocalProblemCont}

To localize the minimization problem, we estimate the discontinuity location as
\begin{equation}
  x_{s,0}=\argmax_{x\in\Omega} |f'(x)|.
  \label{eq:xsInit}
\end{equation}
We then identify local minima of $|f'(x)|$ to the left and right of $x_{s,0}$,
denoted $x_{s,L}$ and $x_{s,R}$, respectively. The localized minimization
problem is solved over $\Omega_{\mathrm{loc}}\equiv[x_{s,L},x_{s,R}]$ as
\begin{subequations}
\begin{equation}
  \bm{\phi}_{\mathrm{loc},*}
  =\argmin_{\bm{\phi}\in\mathcal{X}_{\mathrm{loc}}}
  J_{\mathrm{loc}}(\bm{\phi};f),
\end{equation}
with objective function
\begin{equation}
  \begin{aligned}
  J_{\mathrm{loc}}(\bm{\phi};f)
  &=
  \int_{\Omega_{\mathrm{loc}}}
  \left|
  \mathcal{T}^{-1}(\bm{\phi})(x)-f(x)
  \right|^2 dx\\
  &+
  \lambda_1
  \int_{\Omega_{\mathrm{loc}}}
  \left(
  |f_L'(x)|^2
  +
  |f_R'(x)|^2
  \right) dx\\
  &+\lambda_b(|f_L(x_{s,L})-f(x_{s,L})|^2+|f_R(x_{s,R})-f(x_{s,R})|^2)
  \end{aligned}
\end{equation}
where
\begin{equation}
\mathcal{X}_{\mathrm{loc}}=C^1(\Omega_{\mathrm{loc}})\times
C^1(\Omega_{\mathrm{loc}})\times \Omega_{\mathrm{loc}}\times \mathbb{R}_{>0}.
\end{equation}
\label{eq:1dLSObjLocalCont}
\end{subequations}
The term proportional to $\lambda_b$ penalizes violations of the boundary
values at the ends of the localized region.

Solving~\eqref{eq:1dLSObjLocalCont} gives the localized solution
\begin{equation}
\bm{\phi}_{\mathrm{loc},*}
=
(f_{\mathrm{loc,L,*}},f_{\mathrm{loc,R,*}},x_{s,*},\delta_*).
\end{equation}
Outside $\Omega_{\mathrm{loc}}$, we keep the extension associated with the
physical side equal to the original function and hold the opposite extension
constant at the nearest fitted endpoint as
\begin{subequations}
\begin{align}
  f_{L,*}(x)&=
  \begin{cases}
    f(x) & x<x_{s,L},\\
    f_{\mathrm{loc,L,*}}(x) & x_{s,L}\leq x<x_{s,R},\\
    f_{\mathrm{loc,L,*}}(x_{s,R}) & x_{s,R}\leq x,
  \end{cases}\\
  f_{R,*}(x)&=
  \begin{cases}
    f_{\mathrm{loc,R,*}}(x_{s,L}) & x<x_{s,L},\\
    f_{\mathrm{loc,R,*}}(x) & x_{s,L}\leq x<x_{s,R},\\
    f(x) & x_{s,R}\leq x.
  \end{cases}
\end{align}
\end{subequations}

\subsection{Discrete Formulation of the Local Minimization Problem}
\label{sapp:ls1dDetailsLocalProblemDisc}

In practice, we apply the preceding construction to values of $f$ on a finite
grid. Let $x_i\in\Omega$, $i=1,\dots,N_x$, denote a uniform grid of cell centers
with spacing $\Delta x$, and let $\bm{f}\in\mathbb{R}^{N_x}$ denote the
corresponding value at the cell centers. We approximate the derivative using a
finite-difference matrix $D\in\mathbb{R}^{N_x\times N_x}$, so that $D\bm{f}$
approximates $f'$ on the grid. The discrete estimate of the discontinuity
location is
\begin{equation}
  i_{s,0}=\argmax_{i=1,\dots,N_x}|[D\bm{f}]_i|,
  \qquad
  x_{s,0}=x_{i_{s,0}}.
  \label{eq:xsInitDisc}
\end{equation}
We then identify indices $i_{s,L}<i_{s,0}$ and $i_{s,R}>i_{s,0}$ corresponding
to local minima of $|D\bm{f}|$ to the left and right of $i_{s,0}$, respectively.
This defines the local index set
\begin{equation}
  \mathcal{I}_{\mathrm{loc}}
  =
  \{i_{s,L},i_{s,L}+1,\dots,i_{s,R}\},
\end{equation}
with
\begin{equation}
  \bm{\phi}_h
  =
  (\bm{f}_L,\bm{f}_R,x_s,\delta)
  \in
  \mathbb{R}^{N_{\mathrm{loc}}}
  \times
  \mathbb{R}^{N_{\mathrm{loc}}}
  \times
  [x_{i_{s,L}},x_{i_{s,R}}]
  \times
  \mathbb{R}_{>0},
\end{equation}
where $N_{\mathrm{loc}}=|\mathcal{I}_{\mathrm{loc}}|$ and
$\bm{f}_L,\bm{f}_R$ denote the local grid values of the left and right state
extensions. For $i\in\mathcal{I}_{\mathrm{loc}}$, the discrete reconstruction is
\begin{equation}
  \left[\mathcal{T}_h^{-1}(\bm{\phi}_h)\right]_i
  =
  \frac{f_{L,i}+f_{R,i}}{2}
  -
  \frac{f_{L,i}-f_{R,i}}{2}
  \tanh\left({\frac{x_i-x_s}{\delta}}\right).
  \label{eq:tanh_ls_disc}
\end{equation}

We approximate the continuous objective~\eqref{eq:1dLSObjLocalCont} by
\begin{subequations}
\begin{equation}
  \bm{\phi}_{h,*}
  =
  \argmin_{\bm{\phi}_h\in\mathcal{X}_{\mathrm{loc},h}}
  J_{\mathrm{loc},h}(\bm{\phi}_h;\bm{f}),
\end{equation}
with
\begin{equation}
  \begin{aligned}
  J_{\mathrm{loc},h}(\bm{\phi}_h;\bm{f})
  &=
  \Delta x
  \sum_{i\in\mathcal{I}_{\mathrm{loc}}}
  \left|
  \left[\mathcal{T}_h^{-1}(\bm{\phi}_h)\right]_i-f_i
  \right|^2\\
  &+
  \lambda_1\Delta x
  \left(
  \|D_{\mathrm{loc}}\bm{f}_L\|_2^2
  +
  \|D_{\mathrm{loc}}\bm{f}_R\|_2^2
  \right)\\
  &+
  \lambda_b\left(
  |f_{i_{s,L}}-f_{L,i_{s,L}}|^2
  +
  |f_{i_{s,R}}-f_{R,i_{s,R}}|^2
  \right),
  \end{aligned}
\end{equation}
where
\begin{equation}
  \mathcal{X}_{\mathrm{loc},h}
  =
  \mathbb{R}^{N_{\mathrm{loc}}}
  \times
  \mathbb{R}^{N_{\mathrm{loc}}}
  \times
  [x_{i_{s,L}},x_{i_{s,R}}]
  \times
  \mathbb{R}_{>0}.
\end{equation}
\label{eq:1dLSObjLocalDisc}
\end{subequations}
Here $D_{\mathrm{loc}}$ is the finite-difference matrix restricted to the local
grid. Note that boundary differences inherit the global stencil and can be
computed from the global solution. Further note that, for notational
convenience, entries of local vectors are indexed by their corresponding global
grid indices.

We solve~\eqref{eq:1dLSObjLocalDisc} a bound-constrained trust-region reflective
nonlinear least-squares method, with the Jacobian approximated using two-point
finite differences (see Subsubsection~\ref{sssec:rop}). We initialize the solver
with $\delta_0=\Delta x$, $x_{s,0}$ computed using~\eqref{eq:xsInitDisc}, and
\begin{equation}
  f_{L,0,i}=
  \begin{cases}
    f_i & i \leq i_{s,L},\\
    f_{i_{s,L}} & i > i_{s,L},
  \end{cases}
  \quad
  f_{R,0,i}=
  \begin{cases}
    f_{i_{s,R}} & i < i_{s,R},\\
    f_i & i \geq i_{s,R},
  \end{cases}
  \label{eq:ExtensionsDisc}
\end{equation}
for $i\in\mathcal{I}_{\mathrm{loc}}$. We summarize this initialization procedure
in Algorithm~\ref{alg:1dInit}.

\begin{algorithm}[H]
\caption{Initialization of the 1D Level Set Representation}
\label{alg:1dInit}
\KwIn{$\bm{x}$ and $\bm{f}$}
\KwOut{Initial local extensions $\bm{f}_{L,0}$, $\bm{f}_{R,0}$, discontinuity
estimate $x_{s,0}$, endpoint indices $i_{s,L}$ and $i_{s,R}$, and local index
set $\mathcal{I}_{\mathrm{loc}}$.}
\BlankLine
Compute $D\bm{f}$\;
Compute $i_{s,0}$ and $x_{s,0}$ using~\eqref{eq:xsInitDisc}\;
Identify $i_{s,L}<i_{s,0}$ and $i_{s,R}>i_{s,0}$ as local minima of
$|D\bm{f}|$\;
Set $\mathcal{I}_{\mathrm{loc}}=\{i_{s,L},i_{s,L}+1,\dots,i_{s,R}\}$\;
Compute $\bm{f}_{L,0}$ and $\bm{f}_{R,0}$ using~\eqref{eq:ExtensionsDisc}\;
\end{algorithm}

Solving~\eqref{eq:1dLSObjLocalDisc} gives $\bm{\phi}_{h,*}
=(\bm{f}_{\mathrm{loc},L,*},\bm{f}_{\mathrm{loc},R,*},x_{s,*},\delta_*)$ on the
local grid. As in the continuous setting, outside the local interval the
physical-side extension is set equal to the original grid data and the
opposite extension is held constant so that
\begin{subequations}
\begin{align}
  f_{L,*,i}
  &=
  \begin{cases}
    f_i & i<i_{s,L},\\
    f_{\mathrm{loc},L,*,i} & i_{s,L}\leq i\leq i_{s,R},\\
    f_{\mathrm{loc},L,*,i_{s,R}} & i>i_{s,R},
  \end{cases}\\
  f_{R,*,i}
  &=
  \begin{cases}
    f_{\mathrm{loc},R,*,i_{s,L}} & i<i_{s,L},\\
    f_{\mathrm{loc},R,*,i} & i_{s,L}\leq i\leq i_{s,R},\\
    f_i & i>i_{s,R}.
  \end{cases}
\end{align}
\label{eq:ExtensionsFullDisc}
\end{subequations}
The full discrete level set construction is summarized in
Algorithm~\ref{alg:1dLevelSetMap}.

\begin{algorithm}[H]
\caption{Optimization of the 1D Level Set Representation}
\label{alg:1dLevelSetMap}
\KwIn{$\bm{x}$, $\bm{f}$, $\lambda_1$, and $\lambda_b$}
\KwOut{Full-grid level set variables $\bm{f}_{L,*}$, $\bm{f}_{R,*}$, $x_{s,*}$,
and $\delta_*$.}
\BlankLine
Run Algorithm~\ref{alg:1dInit} to obtain $\bm{f}_{L,0}$, $\bm{f}_{R,0}$,
$x_{s,0}$, $i_{s,L}$, $i_{s,R}$, and $\mathcal{I}_{\mathrm{loc}}$\;
Set the initial width $\delta_0=\Delta x$\;
Form $\bm{\phi}_{h,0}=(\bm{f}_{L,0},\bm{f}_{R,0},x_{s,0},\delta_0)$ on
$\mathcal{I}_{\mathrm{loc}}$\;
Solve~\eqref{eq:1dLSObjLocalDisc}, initialized with $\bm{\phi}_{h,0}$, to obtain
$\bm{\phi}_{h,*}=(\bm{f}_{\mathrm{loc},L,*},
\bm{f}_{\mathrm{loc},R,*},x_{s,*},\delta_*)$\;
Extend $\bm{f}_{\mathrm{loc},L,*}$ and $\bm{f}_{\mathrm{loc},R,*}$ to the full
grid using~\eqref{eq:ExtensionsFullDisc}\;
\end{algorithm}

\begin{remark}
  After solving the optimization problem~\eqref{eq:1dLSObjLocalDisc}, we can
  optionally enforce monotonicity of $\bm{f}_{\mathrm{loc},L}$ and
  $\bm{f}_{\mathrm{loc},R}$ to suppress spurious local oscillations in the
  reconstructed states.
\end{remark}

\subsection{1D Level Set Representation for a Scalar Field with Multiple
Discontinuities: Implementation Details}\label{sapp:1dLSDetails-K}

Here we extend the 1D level set representation methodology for scalar functions
with a single discontinuity to 1D scalar functions with multiple
discontinuities. \ref{sapp:ls1dDetailsLocalProblem} argues that we can fit the
reconstruction map~\eqref{eq:tanh_ls} by solving~\eqref{eq:1dLSObjCont} as a
local, rather than a global, minimization problem because the hyperbolic tangent
function transitions between its extreme values so quickly. Making a similar
argument, we can solve independent local minimization problems to fit the
reconstruction map for solutions with multiple discontinuities.

First, we estimate the center of each discontinuity from local maxima of
$|D\bm{f}|$. For each estimated center, we choose a local neighborhood bounded
by nearby local minima of $|D\bm{f}|$, truncating the neighborhood if necessary
so that it contains no other discontinuity center. We then solve an independent
local minimization problem for each discontinuity, as outlined in
Algorithm~\ref{alg:multiDiscontinuityLocalFit}. When neighboring discontinuities
are too close to admit non-overlapping local neighborhoods, the corresponding
events can instead be fit simultaneously by solving a joint local optimization
problem.

\begin{algorithm}[H]
\caption{Local fitting of the level set representation with multiple discontinuities}
\label{alg:multiDiscontinuityLocalFit}
\KwIn{$\bm{x}$ and $\bm{f}$, number of discontinuities $K$,
regularization parameters $\lambda_1$ and $\lambda_b$.}
\KwOut{Level set variables $\{\bm{f}_{L,k,*},\bm{f}_{R,k,*},
x_{s,k,*},\delta_{k,*}\}_{k=1}^{K}$.}
\BlankLine
Compute $D\bm{f}$\;
Identify $K$ local maxima of $|D\bm{f}|$ with indices ${i}_{s,1},
\dots,{i}_{s,K}$\;
Sort the indices so that $x_{{i}_{s,1}} < x_{{i}_{s,2}} < \cdots
 < x_{{i}_{s,K}}$\;
\For{$k=1,\dots,K$}{
    Select a local neighborhood containing ${i}_{s,k}$ and no other
    discontinuity centers\;
    Restrict $\bm{f}$ and $\bm{x}$ to this neighborhood\;
    Run Algorithm~\ref{alg:1dInit} on the restricted data to obtain
    ${\bm{f}}_{L,k,0}$, ${\bm{f}}_{R,k,0}$, ${x}_{s,k,0}$,
    ${i}_{s,k,L}$, ${i}_{s,k,R}$, and $\mathcal{I}_{\mathrm{loc},k}$\;
    Set ${\delta}_{k,0}=\Delta x$\;
    Form $\bm{\phi}_{h,k,0}=({\bm{f}}_{L,k,0},{\bm{f}}_{R,k,0},
    {x}_{s,k,0},{\delta}_{k,0})$ on $\mathcal{I}_{\mathrm{loc},k}$\;
    Solve~\eqref{eq:1dLSObjLocalDisc}, initialized with $\bm{\phi}_{h,k,0}$, to
    obtain $\bm{\phi}_{h,k,*}=({\bm{f}}_{\mathrm{loc},k,L,*},
    {\bm{f}}_{\mathrm{loc},k,R,*},{x}_{s,k,*},{\delta}_{k,*})$\;
    Extend ${\bm{f}}_{\mathrm{loc},k,L,*}$ and
    ${\bm{f}}_{\mathrm{loc},k,R,*}$ to the full grid
    using~\eqref{eq:ExtensionsFullDisc}\;
    Store $({\bm{f}}_{L,k,*},{\bm{f}}_{R,k,*},
    {x}_{s,k,*},{\delta}_{k,*})$\;
}
\end{algorithm}

\subsection{Level Set Representation for 1D Compressible Euler Equations}\label{sapp:1d-comp}

Here we extend the level set representation for scalar functions to 
vector-valued functions, such as those obtained from solutions to the 1D Euler
equations. For a single flow field $\bm{w}=(\rho,u,p)$, we apply the level set
representation method field-by-field, as shown in
Algorithm~\ref{alg:field-by-field-primitive}.

To avoid updating the shock width during the EnKF
analysis step, we fit level set representations such that all ensemble members
use the same event widths. First, we fit a level set representation to each
ensemble member separately, and then to refit the level set representation for
all ensemble members using the median shock width, as outlined in
Algorithm~\ref{alg:field-by-field-primitive-ens}.

\begin{algorithm}
\caption{Level set representation for a single 1D flow}
\label{alg:field-by-field-primitive}
\KwIn{Primitive fields $\bm{w}=(\rho,u,p)$, expected discontinuity
counts $\bm{K}$, and optionally prescribed shock widths
$\bm{\delta}_{\mathrm{c}}=(
\delta_{\rho,\mathrm{c}},
\delta_{u,\mathrm{c}},
\delta_{p,\mathrm{c}})$.}
\KwOut{Level set representation}
\BlankLine
\For{each field $w_i$ with $K_i>0$}{
    \If{$\delta_{\mathrm{c},i}$ is not prescribed}{
        Apply Algorithm~\ref{alg:multiDiscontinuityLocalFit} to $w_i$\;
        
    }
    \Else{
      Apply Algorithm~\ref{alg:multiDiscontinuityLocalFit} to $w_i$ with
      prescribed $\delta_{\mathrm{c},i}$\;
    }
}
\end{algorithm}

\begin{algorithm}
\caption{Level set representation for an ensemble of 1D flows}
\label{alg:field-by-field-primitive-ens}
\KwIn{Primitive-field ensemble $\{\bm{w}^{(n)}\}_{n=1}^{N_e}$
and expected discontinuity counts $\bm{K}$.}
\KwOut{Ensemble of level set representations, fitted with common widths.}
\BlankLine
\For{$n=1,\dots,N_e$}{
    Apply Algorithm~\ref{alg:field-by-field-primitive} to
    $\bm{w}^{(n)}$ without prescribed widths and obtain
    $\bm{\delta}^{(n)}$\;
}
\For{each field $w_i$ with $K_i>0$}{
    Set $\delta_{c,i}
    =\operatorname{median}_{1\leq n\leq N_e}\delta_i^{(n)}$\;
}
\For{$n=1,\dots,N_e$}{
    Apply Algorithm~\ref{alg:field-by-field-primitive} to
    $\bm{w}^{(n)}$ with prescribed field widths $\bm{\delta}_c$\;
}
\end{algorithm}

\subsection{Affine Alignment of Smoothly Varying Features}
\label{ssec:ls1d-affine}

The hyperbolic tangent function can be used to model sharp features. However,
it is not suitable to model features that are slowly and continuously varying,
such as a rarefaction wave in a compressible flow. Although rarefaction waves
vary continuously, and slowly compared to shocks and contact surfaces, we
observe that the standard EnKF still performs poorly for these features.
Therefore, we additionally handle rarefaction waves by aligning them using
an affine transformation in $x$.

Let us denote this transformation by
\begin{subequations}
\begin{equation}
\mathcal{R}_{(a,b)}(f)(x)=f(ax+b),
\end{equation}
with inverse
\begin{equation}
\mathcal{R}_{(a,b)}^{-1}(f)(x)=f((x-b)/a),
\end{equation}
\label{eq:affine1d}
\end{subequations}
where we restrict to $a>0$ so that the spatial map is invertible and orientation
preserving. To align $f^{(2)}(x)$ with $f^{(1)}(x)$, we compute
\begin{subequations}
\begin{equation}
    s^{(1)}(x)=\big|\frac{df^{(1)}}{dx}\big|,\quad
    s^{(2)}(x)=\big|\frac{df^{(2)}}{dx}\big|,
\end{equation}
which we normalize as
\begin{equation}
    {s}^{(1)*}(x)=\frac{s^{(1)}(x)}{\max_{x\in\mathcal{D}}s^{(1)}(x)},\quad
    {s}^{(2)*}(x)=\frac{s^{(2)}(x)}{\max_{x\in\mathcal{D}}s^{(2)}(x)}.
\end{equation}
Then we solve the minimization problem
\begin{equation}
    (a,b)=\argmin_{a\in\mathbb{R}_{>0},b\in\mathbb{R}} 
    \|
    {s}^{(1)*}(x)-\mathcal{R}_{(a,b)}({s}^{(2)*})
    \|_2^2,
\end{equation}
\end{subequations}
which minimizes the difference in their normalized gradients. This affine
transformation is similar to the registration procedure employed by the morphing
EnKF~\cite{beezley2008registration}, but it comprises only two parameters, and
is thus much simpler.

\section{Implementation Details for 2D Level Set Representations}\label{app:ls2dDetails}

This Appendix presents details on identifying the level set representation for
2D compressible flows. \ref{sapp:ls2dDetails-Zero} describes how to identify the
zero level set for a discontinuity in 2D, and how to subsequently construct the
level set function.

\subsection{Identifying the Zero Level Set
of a 2D Scalar Field}\label{sapp:ls2dDetails-Zero}

For 1D scalar field, we obtain an initial estimate of the discontinuity
location by identifying local maxima of the solution gradient, as described
in \ref{sapp:ls1dDetailsLocalProblemCont}. This subsection extends the procedure
to a 2D scalar field.

Consider the domain $\Omega\subset\mathbb{R}^2$ with coordinate
$\bm{x}\in\mathbb{R}^2$. Let $f:\Omega\to\mathbb{R}$ denote a 2D scalar field,
and let us assume that there exists a discontinuity of $f$ in $\Omega$. Let
$\bm{X}(s)$ denote a set of points along this discontinuity, parameterized by arc
length $s$. For closed curves, $\bm{X}(s)$ is periodic with period $L$, and we
need only consider $s\in[0,L)$. From this, we obtain the curve
\begin{equation}
    \Gamma\equiv\{\bm{X}(s):s\in[0,L)\},
    \label{eq:ls2Dcurve}
\end{equation}
The gradient of $f$
\begin{equation}
    \nabla f=(\partial_x f,\partial_y f),
\end{equation}
and its norm
\begin{equation}
    s_1(\bm{x})=\|\nabla f(\bm{x})\|_2,
\end{equation}
will be large near $\Gamma$. Thus, we identify the point
\begin{equation}
    {\bm{x}}_{s,0}
    =
    \argmax_{\bm{x}\in\Omega}
    s_1(\bm{x}),
\end{equation}
where the gradient norm is maximal, and we take this to be the first point in
our discontinuity curve, so that
\begin{equation}
    \Gamma_0\equiv\{x_{s,0}\}.
\end{equation}
The gradient of $f$ is locally normal the discontinuity curve, so the local
unit tangent vector is
\begin{equation}
    \bm{t}(\bm{x})
    =
    \frac{1}{\|\nabla f(\bm{x})\|_2}
    \begin{bmatrix}
        -\partial_y f(\bm{x})\\
        \partial_x f(\bm{x})
    \end{bmatrix}.
\end{equation}
We seek the next point on the discontinuity curve by maximizing the gradient
magnitude along the tangent direction $\bm{t}$. Given the current point
${\bm{x}}_{s,i}$, we compute
\begin{equation}
    {\bm{x}}_{s,i+1}
    =
    \argmax_{\bm{x}\in\{{\bm{x}}_{s,i}+r\bm{t}({\bm{x}}_{s,i})
    :r\in(0,\Delta r]\}}s_1(\bm{x}),
\end{equation}
for small $\Delta r>0$.
In practice, we approximate this search by selecting the adjacent grid point
with the largest gradient magnitude in the quadrant indicated by the local
tangent to the current grid point. The step length is therefore at most one grid
diagonal, and we do not prescribe $\Delta r$. Repeating this procedure until the
tracing terminates gives a curve that parameterizes the discontinuity location
\begin{equation}
    {\Gamma}_N
    =
    \{{\bm{x}}_{s,i}\}_{i=0}^{N}.
\end{equation}
Tracing terminates when the next selected grid point has already been visited,
which is not necessarily the starting point.

\subsection{Identifying Smooth State Extensions for a Scalar Field}\label{sapp:ls2dRep}

For 2D scalar fields, we obtain smooth, globally defined, state extensions
by applying the 1D optimization methodology described in
\ref{sapp:ls1dDetailsLocalProblemDisc} locally normal to the curve
${\Gamma}$. Given a point ${\bm{x}}_{s,i}\in{\Gamma}$, we compute the normal
vector
\begin{equation}
    \bm{n}({\bm{x}}_{s,i})
    =
    \frac{1}{\|\nabla f({\bm{x}}_{s,i})\|_2}
    \begin{bmatrix}
        \partial_x f({\bm{x}}_{s,i})\\
        \partial_y f({\bm{x}}_{s,i})        
    \end{bmatrix}.
\end{equation}
We then define the line passing through ${\bm{x}}_{s,i}$ in the normal
direction,
\begin{equation}
    \bm{l}_i(\eta)
    =
    {\bm{x}}_{s,i}
    +
    \eta\bm{n}({\bm{x}}_{s,i}),
\end{equation}
where $\eta\in\mathbb{R}$ denotes the signed distance from ${\bm{x}}_{s,i}$
along the normal direction. Restricting the solution to this line yields the
1D profile
\begin{equation}
    f_i(\eta)
    =
    f
    (
    {\bm{x}}_{s,i}
    +
    \eta\bm{n}({\bm{x}}_{s,i})
    ),
\end{equation}
to which we apply the 1D fitting procedure described in
\ref{sapp:ls1dDetailsLocalProblemDisc}. We additionally compute the local level
set function along $\bm{l}_i$ as
\begin{equation}
    \varphi_i(\eta)=\eta.
\end{equation}
Repeating this procedure for every $i=1,\dots,N$ yields a set of curves locally
normal to the discontinuity. We compute global functions by interpolating using
radial basis functions.

\subsection{Level Set Representation for the 2D Euler Equations}\label{sapp:ls2dApplyEuler}

To construct a level set representation for a solution to the Euler equations,
we apply the methodologies developed in \ref{sapp:ls2dDetails-Zero} and
\ref{sapp:ls2dRep} directly to the density and pressure fields. For the velocity
field, we identify the zero level set by applying the procedure outlined in
\ref{sapp:ls2dDetails-Zero} to the norm of the velocity field $\|\bm{u}\|$. To
compute the state extensions, we decompose the velocity field into its normal
and tangential components
\begin{equation}
    u_n=\bm{u}\cdot\bm{n},
    \qquad
    u_t=\bm{u}\cdot\bm{t},
\end{equation}
where $\bm{n}$ and $\bm{t}$ denote the unit normal and tangent vectors,
respectively. We then apply the methodology from \ref{sapp:ls2dRep}
to $u_n$, obtaining a level set representation of the normal velocity component.
The level set representations of the Cartesian velocity components are
subsequently recovered using $u_t$ and the level set representation of $u_n$:
\begin{equation}
    u_k = u_{n,k} n_x - u_t n_y,
    \qquad
    v_k = u_{n,k} n_y + u_t n_x,
    \qquad
    k=1,\ldots,N_d+1,
\end{equation}
where $N_d$ denotes the number of discontinuities. Once we have fit the state
extensions to $u_n$ and $u_t$, we recover $u$ and $v$ using the above relations.

\subsection{2D Registration Process}\label{sapp:ls2Dregistration}

This subsection introduces the 2D registration process. First we align the
level set functions. Then, using information from this alignment, we align the
state extensions corresponding to the rarefaction wave.
\ref{ssapp:ls2DregistrationLS} introduces the procedure to align the level set
functions, while \ref{ssapp:ls2DregistrationLF} introduces the procedure for the
state extensions.

\subsubsection{2D Registration Process: Level Set Functions}\label{ssapp:ls2DregistrationLS}

Solutions to the 2D blast wave problem produce zero level sets that are
circular, so we introduce the registration map
\begin{equation}
    \mathcal{R}^{(2D)}_{(a^{(i)},b^{(i)},\bm{c}^{(i)},\bm{c}^{(1)})}
    (\varphi^{(i)})(\bm{x})
    = \varphi^{(i)}\left(
    \bm{c}^{(i)}
    +
    \frac{\|\bm{x}-\bm{c}^{(1)}\|-b^{(i)}}
    {a^{(i)}\|\bm{x}-\bm{c}^{(1)}\|}
    \left(\bm{x}-\bm{c}^{(1)}\right)
    \right)
    \label{eq:align2D}
\end{equation}
which aligns the center of the zero level set of $\varphi^{(i)}$ to that of
$\varphi^{(1)}$, while also scaling and shifting the radial coordinate. Here,
$\bm{c}^{(i)}$ denotes the center of the zero level set for $\varphi^{(i)}$,
while $a^{(i)}$ is a radial scaling parameter, and $b^{(i)}$ is a radial shift.
We need only align the centers of the level set functions, so we set $a^{(i)}=1$
and $b^{(i)}=0$. We independently fit centers to the zero level set of each
level set function, and subsequently evaluate the registration map
$\mathcal{R}^{(2D)}_{(1,0,\bm{c}^{(i)},\bm{c}^{(1)})}(\varphi^{(i)})$ for
$i=2,\dots,N_e$.

To estimate the center of a circle, we first note that, if a circle is centered
at $\bm{c}^{(i)}=(c^{(i)}_x,c^{(i)}_y)$ with radius $r^{(i)}$, then
\begin{subequations}
\begin{equation}
(x-c^{(i)}_x)^2
+(y-c^{(i)}_y)^2
=(r^{(i)})^2,
\end{equation}
which can be rewritten as
\begin{equation}
x^2+y^2+dx+ey+f=0,
\label{eq:ls2dCirc}
\end{equation}
for
\begin{equation}
d=-2c^{(i)}_x,\quad
e=-2c^{(i)}_y,\quad
f=(c^{(i)}_x)^2+(c^{(i)}_y)^2+(r^{(i)})^2.
\end{equation}
We can then use~\ref{eq:ls2dCirc} to fit $d$, $e$, and $f$ using linear least
squares, and we recover the desired parameters as
\begin{equation}
c^{(i)}_x=-\frac{d}{2},\quad
c^{(i)}_y=-\frac{e}{2},\quad
r^{(i)}=\sqrt{\frac{d^2+e^2}{2^2}-f}.
\end{equation}
\end{subequations}
We then use~\eqref{eq:align2D} with $a^{(i)}=1$ and $b^{(i)}=0$ to align the
level set functions

\subsubsection{Aligning the 2D State Extensions}\label{ssapp:ls2DregistrationLF}

Given the $\bm{c}^{(i)}$ from the level set function $\bm{\phi}^{(i)}$,
corresponding, to the density field $\rho^{(i)}$, we align the density field
$\rho^{(i)}$ to the density field $\rho^{(1)}$ by seeking $a^{(i)}$ and
$b^{(i)}$ that minimizes the difference between the gradients of their
solutions. We compute the solution gradients as
\begin{equation}
    s^{(i)}(\bm{x})=
    \sqrt{
    \frac{\partial\rho^{(i)}}{\partial x}^2
    +
    \frac{\partial\rho^{(i)}}{\partial y}^2
    }
\end{equation}
and normalize them As
\begin{equation}
    {s}^{(i)*}(\bm{x})=
    \frac{s^{(i)}(\bm{x})}{\max_{\bm{x}\in\mathcal{D}}s^{(i)}(\bm{x})}.
\end{equation}
Then we solve the minimization problem
\begin{equation}
    (a^{(i)},b^{(i)})=\argmin_{a\in\mathbb{R}_{>0},b\in\mathbb{R}} 
    \|
    {s}^{(1)*}(\bm{x})-\mathcal{R}^{(\mathrm{2D})}_{(a,b,\bm{c}^{(i)},\bm{c}^{(1)})}({s}^{(i)*})(\bm{x})
    \|_2^2.
\end{equation}
This mimics the 1D procedure developed in \ref{ssec:ls1d-affine}, but applied
radially.

\end{document}